\documentclass[usegraphicx,useAMS,usenatbib,]{mn2e}
\usepackage{aas_macros}

\newcommand{\lori}{\mbox{$\lambda$~Ori}}
\newcommand{\sori}{\mbox{$\sigma$~Ori}}
\newcommand{\kms}{\mbox{${\rm km\,s}^{-1}$}}
\newcommand{\Msolar}{\mbox{${M}_{\sun}$}}

\newcommand{\Vr}{\mbox{${\rm V}_{\rm r}$}}
\newcommand{\Rc}{\mbox{${\rm R}_{\rm C}$}}
\newcommand{\Ic}{\mbox{${\rm I}_{\rm C}$}}
\newcommand{\Vbar}{\mbox{$\bar{{\rm V}}_{\rm r}$}}
\newcommand{\Vsini}{\mbox{${\rm V}_{\rm rot}\sin i$}}
\newcommand{\logg}{\mbox{$\log\,{\rm g}$}}
\newcommand{\ChaHa}{\mbox{Cha~H$\alpha$~8}}

\title[Spectroscopic binaries near \sori\ and  \lori.]
  {A survey for low mass spectroscopic binary stars in the young clusters
around $\sigma$ Orionis and $\lambda$ Orionis.}
\author[P.F.L. Maxted et~al.]
  {P.F.L.~Maxted$^1$, R.D.~Jeffries$^1$, J.M.~Oliveira$^1$, T.~Naylor$^2$,
 R.J.~Jackson$^1$ \\
  $^1$Astrophysics Group,  Keele University, Keele, 
      Staffordshire ST5 5BG, United Kingdom\\
$^2$School of Physics, University of Exeter, Stocker Road, Exeter, EX4~4QL
}
\date{Submitted 2007}

\pagerange{\pageref{firstpage}--\pageref{lastpage}} \pubyear{2006}

\def\LaTeX{L\kern-.36em\raise.3ex\hbox{a}\kern-.15em
    T\kern-.1667em\lower.7ex\hbox{E}\kern-.125emX}

\newcommand{\Ntarget}{218}
\newcommand{\Nmember}{196}

\newcommand{\Nvariable}{12}
\newcommand{\Nbinmem}{12}
\newcommand{\Nrvvmem}{11}

\begin{document}
\label{firstpage}

\maketitle

\begin{abstract}
 We have obtained multi-epoch, high-resolution spectroscopy of \Ntarget\
candidate low-mass stars and brown dwarfs in the young clusters around \sori\
and \lori. We find that \Nmember\ targets are cluster members based on their
radial velocity, the equivalent width of their Na\,I\,8200 lines and the
spectral type from their TiO band strength. We have identified \Nrvvmem\ new
binary stars among the cluster members based on their variable radial velocity
and an additional binary from the variation in its line width and shape. Of
these, 6 are double-lined spectroscopic binaries (SB2) where the components of
the binary are of comparable brightness. The others are single-lined binaries
(SB1) in which the companion is faint or the spectra of the stars are blended.
There are  3 narrow-lined SB1 binaries in our sample for which the companion
is more than 2.5 magnitudes fainter than the primary. This suggests that the
mass ratio distribution for the spectroscopic binaries in our sample is broad
but that there may be a peak in the distribution near $q=1$. The sample covers
the magnitude range $\Ic = 14$\,--\,18.9 (mass $\approx 0.55 \--
0.03\Msolar$), but all of the binary stars are brighter than $\Ic=16.6 $ (mass
$\approx 0.12\Msolar$) and 10 are brighter than $\Ic = 15.5$ (mass $\approx
0.23\Msolar$). There is a significant lack of spectroscopic binaries in our
sample at faint magnitudes even when we account for the decrease in
sensitivity with increasing magnitude. We can reject the hypothesis that the
fraction of spectroscopic binaries is a uniform function of \Ic\ magnitude
with more than 99 percent confidence. The spectroscopic binary fraction for
stars more massive than about $0.1\Msolar$ ($\Ic < 16.9$) is $f_{\rm bright} =
0.095^{+0.012}_{-0.028}$. The 90 percent confidence upper limit to the
spectroscopic binary fraction for very low mass (VLM) stars (mass $<
0.1\Msolar$) and brown dwarfs (BDs) is $f_{\rm faint} < 7.5$\,percent. The
hypothesis that $f_{\rm bright}$ and $f_{\rm faint}$  are equal can be
rejected with 90\,percent confidence.  The average detection probability for
our survey is 50\,percent or more for binaries with separations  up to
0.28\,au for stars with $\Ic < 16.9$ and 0.033\,au for the fainter stars in
our sample. We conclude that we have found strong evidence for a change in the
fraction of spectroscopic binaries  among young VLM stars and brown dwarfs
when compared to more massive stars in the same star-forming region. This
implies a difference in the total binary fraction between VLM stars and BDs
compared to more massive stars or a difference in the distribution of
semi-major axes, or both.

\end{abstract}


\section{Introduction}
The origins of very low-mass stars (VLMS) and brown dwarfs (BD) are proving
difficult to understand, despite being more common than stars of higher mass.
Ideas include ejection from protostellar aggregates
\citep{2001AJ....122..432R}, formation within convergent flows generated by
turbulence \citep{2004ApJ...617..559P}, the photo-erosion of pre-stellar cores
\citep{2004A&A...427..299W} or fragmentation within the outer parts of
circumstellar discs \citep{2006A&A...458..817W}.

 The frequency and separation distribution of binary systems is an important
constraint on the likely formation process. There is strong evidence that the
binary properties of the lowest mass stars and brown dwarfs are quite
different to those of higher mass objects (see the review of
\citealt{2007prpl.conf..427B}). Resolved imaging of nearby VLMS and BDs in the
field show that about 15--20 percent of systems are binaries with separations
greater than 1--2\,au, but that very few have separations greater than 20\,au
(e.g. \citealt{2003ApJ...587..407C}; \citealt{2003AJ....126.1526B}). This
contrasts with higher mass stars where overall binary frequencies are 30--60
percent, with a much broader spread of possible separations
(\citealt{1992ApJ...396..178F}, \citealt{1991A&A...248..485D}).

 A missing part of the picture is how many VLMS and BDs are binary systems
with separations less than about 1\,au, where the imaging observations cannot
reach. \citet{2005MNRAS.362L..45M} used previously published radial velocity
(RV) results to show that an overall binary frequency (at all separations) of
32--45 percent was needed to explain the presence of several binaries detected
by RV variations, with most of them at small separations. This high frequency
found support from a small RV survey by \citet{2006MNRAS.372.1879K} who found
a frequency of 11\,--\,40 percent for separations less than 0.1\,au in a young
association. On the other hand \citet{2006AJ....132..663B} found few RV
variables in their survey of field VLMS/BDs and concluded that the overall
binary frequency was 16--36 percent, with very few binaries at separations
below 1\,au

 The situation is unresolved and it is quite likely that the apparent
discrepancies between these various authors arise from biases within the
samples considered; differences in analysis technique and that in terms of
absolute numbers, very few short-period VLMS/BD binary systems have yet  been
found, thus limiting the statistical precision possible. 

 In this paper we present the results of an RV survey of a large number of
low-mass stars and brown dwarfs in the \sori\ and \lori\ clusters. These
clusters are young and nearby (3--5 Myr, 330--450\,pc) and contain large
populations of VLMS and BDs (\citealt{2004AN....325..705B};
\citealt{2004ApJ...610.1064B}; \citealt{2005MNRAS.356...89K}).  We have
observed more than 200 objects using the {\sc flames} multi-fibre spectrograph
on the VLT-Kueyen telescope to measure radial velocities at several epochs and
searched for close binary systems at a range of masses.

\section{Observations and data reduction}
\label{ObsAndRed}

\subsection{Observations}
\label{Observations}

 We have used the {\sc flames} multi-object spectrograph
\citep{2002Msngr.110....1P} on ESO's VLT UT2 (Kueyen) telescope to obtain
multi-epoch, high-resolution spectroscopy of \Ntarget\ faint stars in the
clusters around \sori\ and \lori. The instrument is capable of providing
spectra of up to 130 targets in one setting over a field of view 25 arcmin in
diameter.

\begin{figure*}
\includegraphics[width=0.47\textwidth]{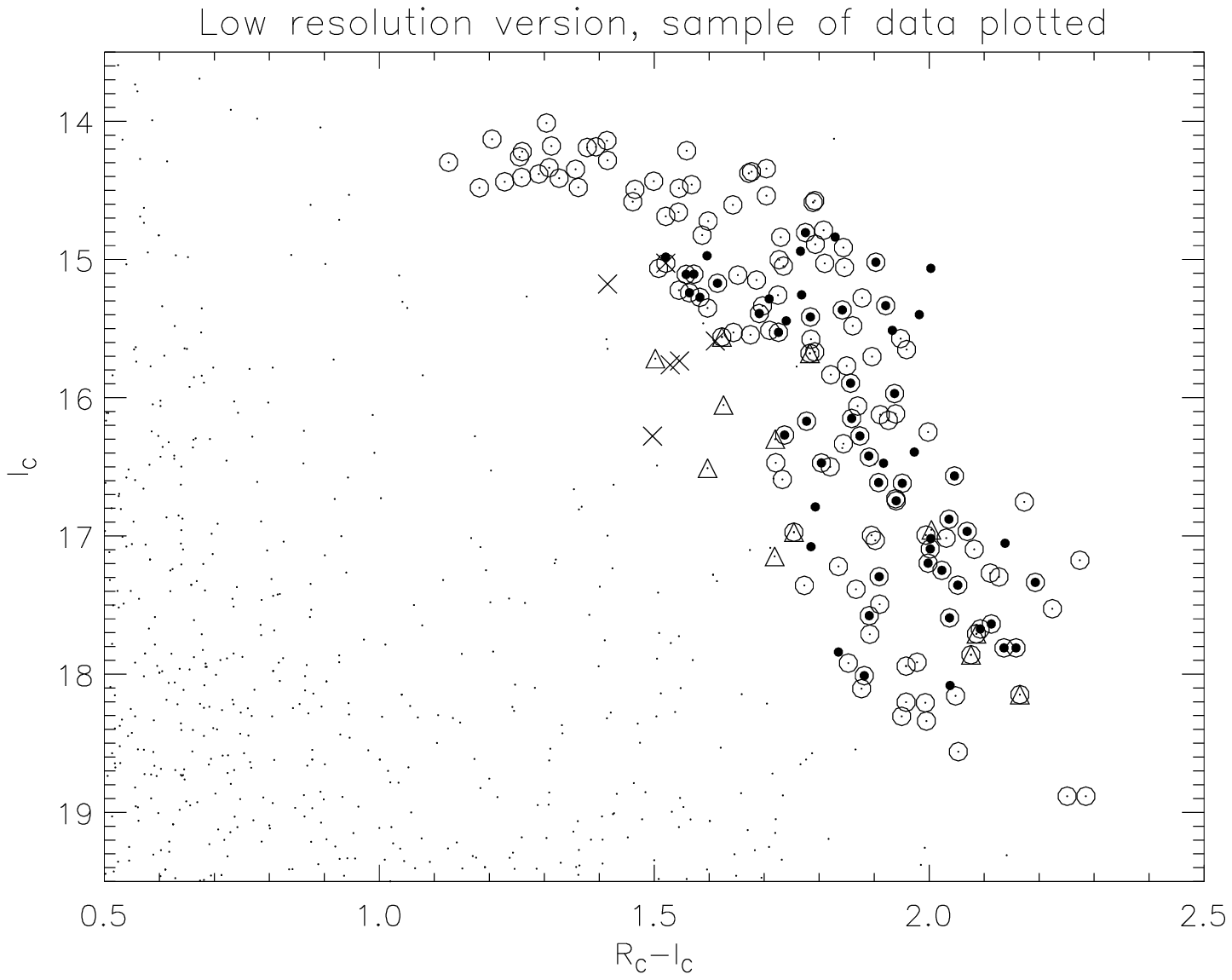}
\includegraphics[width=0.47\textwidth]{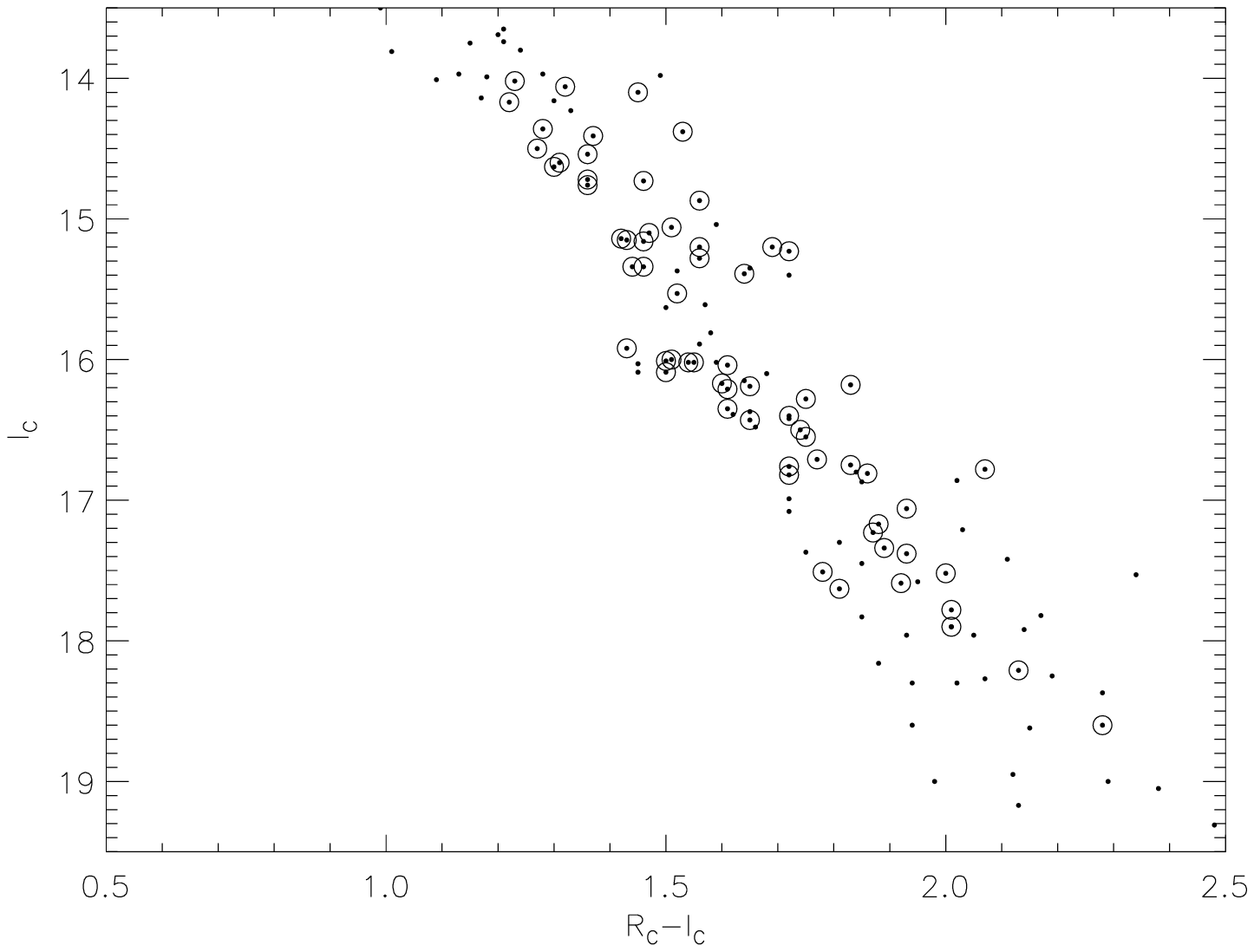}
\caption{Colour\,--\,magnitude diagrams for stars around \sori\ (left panel) 
and  \lori\ (right panel) with the stars we have observed highlighted
(open circles). Other symbols in the left panel indicate the membership status
for stars from \citet{2005MNRAS.356...89K} as follows: filled circles --
member; triangles -- uncertain membership; crosses -- non-member. This
membership status was ignored during the fibre allocation process.
\citet{2004ApJ...610.1064B} only provide data for stars they consider to be
candidate members of the \lori\ cluster so we are not able to show that
position of other stars in the right panel.
\label{RIFig} }
\end{figure*}

 We selected targets around \sori\ from the photometric catalogue of
\citet{2005MNRAS.356...89K}. We included all stars in the magnitude range $14
< \Ic < 19$ with the correct \Ic\ magnitude and $(\Rc - \Ic)$ colour to be a
cluster member in the input catalogue for the fibre allocation process,
irrespective of any other membership information that may have been available.
For \lori\ all the candidate members identified in the catalogue of
\citet{2004ApJ...610.1064B} were included in the input catalogue for the fibre
allocation process. Fibres were allocated in 6 fields, 2 fields near \lori\
and 4 fields near \sori. The central position of each field and the number of
targets for which useful spectra were obtained are given in
Table~\ref{ObsDateTable}. The position of the targets in the \Ic\ v. $(\Rc -
\Ic)$ colour-magnitude diagram is shown in Fig.~\ref{RIFig}. Those fibres that
could not be allocated to targets were used to obtain simultaneous background
spectra for sky subtraction. The two fields around \lori\ overlap so 10 stars
were observed in 2 fields.  For most stars we have obtained 3\,--\,4 spectra
with the exception of 10 stars observed in 2 fields for which we typically
have 6\,--\,7 spectra. The baseline of the observations is at least 24 days
for all stars and is more typically 60\,--\,70 days. The sensitivity of our
survey drops rapidly for binaries with semi-major axes $a\goa 0.25$\,au as a
result of the baseline of our observations.

\begin{figure*}
\includegraphics[width=0.98\textwidth]{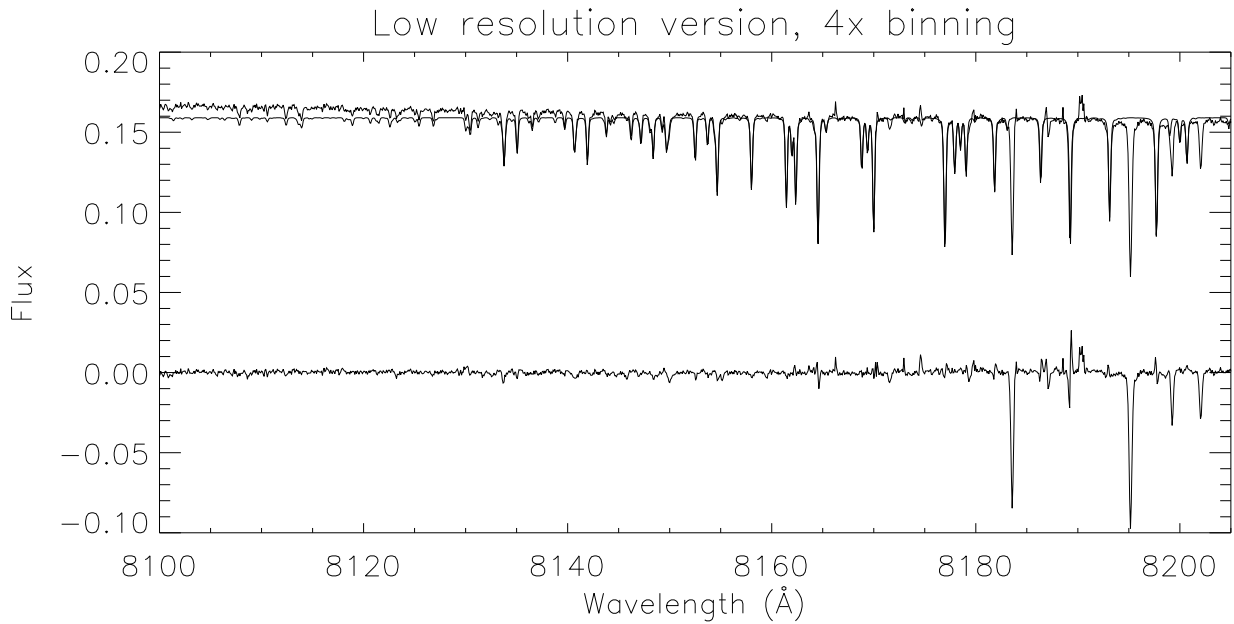}
\caption{\label{TelluricFig}Example of a synthetic telluric absorption
spectrum from a model terrestrial atmosphere fit to the telluric absorption in
a section of one of our {\sc uves} spectra. 
 The residuals from the fit of the synthetic absorption spectrum plus a
low order polynomial are shown below the {\sc uves} spectrum and the synthetic
absorption spectrum. Note that the residuals include absorption lines from the
star, e.g., the Na\,I~8200\AA\ doublet.  }
\end{figure*}

\begin{figure*}
\includegraphics[angle=270,width=0.95\textwidth]{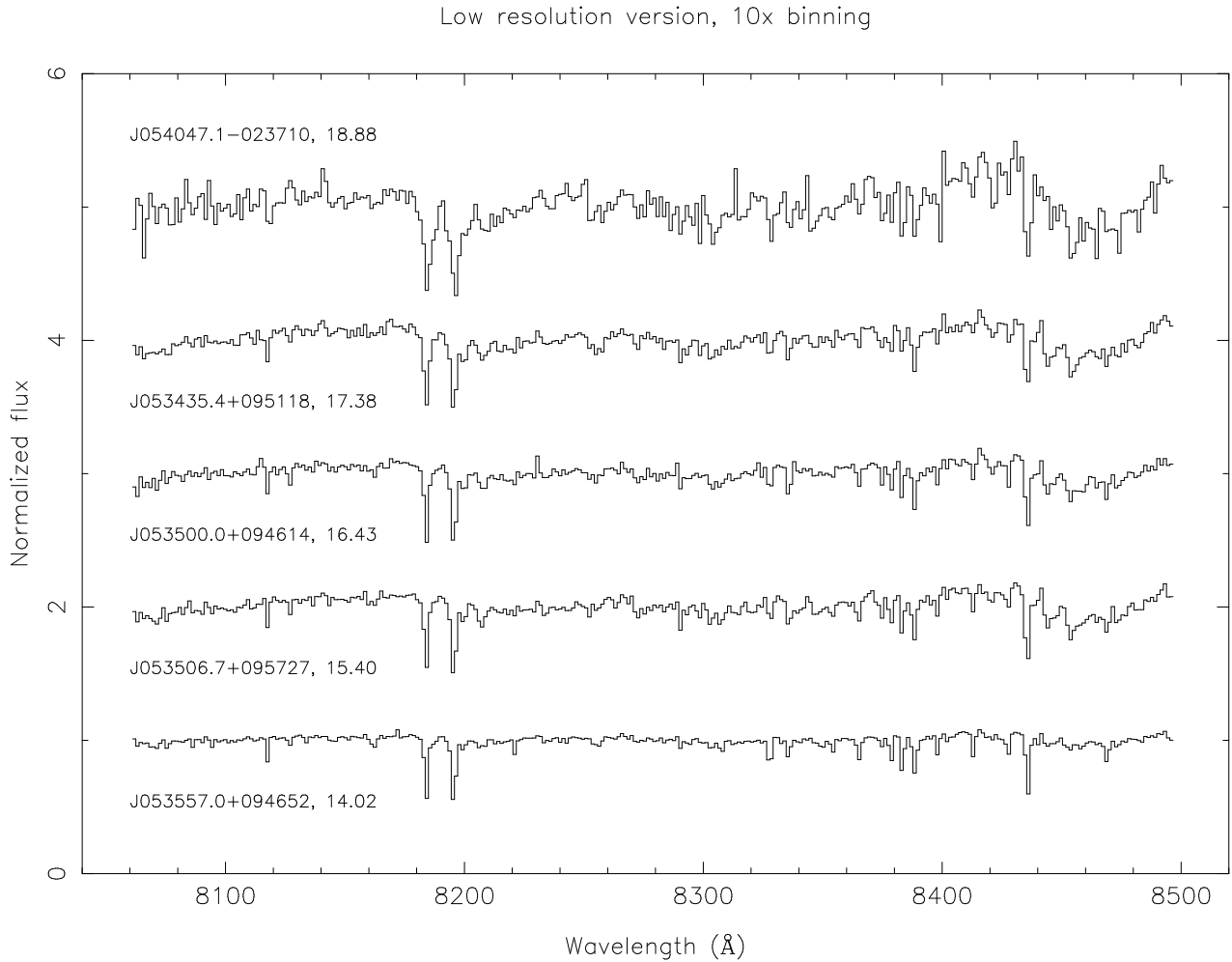}
\caption{\label{ExampleSpectraFig}Example spectra of apparently single stars
from our sample. The spectra of each star have been normalized by a low-order
polynomial and offset for clarity and are labeled by name and I-band
magnitude. 
 } \end{figure*}

 \begin{table}
 \caption{Log of observations. The number of stars observed with the {\sc
giraffe} spectrograph on each date for which useful spectra (signal-to-noise
ratio
$>5$) were obtained is given in the column headed $N$. The date of observation
is given in the fifth column. The standard deviation of the
radial velocities of the night sky emission line spectra obtained at each
epoch, $\sigma_{\rm sky}$, are given in the final column. 
\label{ObsDateTable}}
 \begin{center}
 \begin{tabular}{@{}lrrrrr}
  \multicolumn{1}{@{}l}{Field}  &
  \multicolumn{2}{c}{Position } &
  \multicolumn{1}{l}{$N$} &
  \multicolumn{1}{l}{Date} &
  \multicolumn{1}{l}{$\sigma_{\rm sky}$}\\
 &
  \multicolumn{1}{c}{$\alpha_{\rm  J2000}$} &
  \multicolumn{1}{c}{$\delta_{\rm  J2000}$} &
  & 
  \multicolumn{1}{c}{(UTC)}&
  \multicolumn{1}{c}{(\kms)}\\
 \hline
 \noalign{\smallskip}
 $\lambda$~Ori\,1 & 05\,35\,58.5& $+$09\,51\,21  &  34  &2005\,10\,14 & 0.66\\
                 &             &            &  34  &2005\,11\,17 & 0.59\\
                 &             &            &  35  &2005\,12\,30 & 0.67\\
 $\lambda$~Ori\,2 & 05\,34\,44.1& $+$09\,51\,44  &  49  &2005\,10\,14 & 0.64\\
                 &             &            &  49  &2005\,11\,27 & 0.99\\
                 &             &            &  50  &2005\,11\,28 & 0.63\\
                 &             &            &  50  &2005\,12\,30 & 0.58\\
 $\sigma$~Ori\,1  & 05\,40\,28.0& $-$02\,15\,35  &  26  &2005\,11\,13 & 0.60\\
                 &             &            &  22  &2005\,11\,17 & 0.60\\
                 &             &            &  26  &2006\,01\,08 & 0.64\\
                 &             &            &  26  &2006\,01\,13 & 0.92\\
 $\sigma$~Ori\,2  & 05\,40\,01.6& $-$02\,40\,02  &  17  &2005\,11\,13 & 0.67\\
                 &             &            &  29  &2005\,12\,07 & 0.65\\
                 &             &            &  28  &2006\,01\,13 & 0.74\\
 $\sigma$~Ori\,3  & 05\,38\,26.9& $-$02\,41\,21  &  56  &2005\,11\,13 & 0.64\\
                 &             &            &  55  &2005\,11\,24 & 0.67\\
                 &             &            &  56  &2005\,12\,07 & 0.56\\
                 &             &            &  55  &2006\,01\,13 & 0.60\\
 $\sigma$~Ori\,4  & 05\,38\,23.8& $-$02\,14\,59  &  33  &2005\,11\,11 & 0.59\\
                 &             &            &  34  &2005\,11\,11 & 0.58\\
                 &             &            &  34  &2005\,12\,07 & 0.62\\
                 &             &            &  32  &2006\,01\,13 & 0.60\\
 \noalign{\smallskip}
 \end{tabular}   
 \end{center}    
 \label{resultstable}
 \end{table}

 Light from the fibres was fed to the {\sc giraffe} spectrograph operated in a
high resolution mode with the H836.6 echelle grating. A filter was used to
select light from the 6$^{th}$ echelle order which covers the wavelength
region  8073\,--\,8632\AA. The resolving power of the spectra is $R\approx
16000$. Spectra were obtained in service mode by ESO staff on the dates given
in Table~\ref{ObsDateTable}. The exposure time in each case was 2750\,s. The
seeing during the exposures was typically 0.92\,arcsec but varied from
0.48\,arcsec to 1.45\,arcsec.

 We also took advantage of the possibility to use up to 8 fibres from {\sc
flames} to obtain spectra with the {\sc uves} echelle spectrograph
\citep{2000SPIE.4008..534D} at the same time as the {\sc giraffe}
observations. We used the CD\#4 echelle grating to obtain spectra at a
resolution of $R\approx 47000$ covering the same spectral range as the  {\sc
giraffe} spectra for one or two bright, early-type stars in the field. These
spectra were used to calibrate the telluric absorption in the  {\sc giraffe}
spectra. Fibres not allocated to bright stars were used to obtain simultaneous
spectra of the night sky.

\subsection{Reduction of the spectra}
 There is strong fringing in the images produced by the {\sc giraffe}
spectrograph when operated at the wavelengths we have used for our
observations. This makes the extraction of the spectra problematic. For
brighter stars it is possible to use the spectra and their associated errors
extracted from the data automatically by the ESO pipeline provided by the
observatory. These spectra are extracted from the images by summing the pixels
in each row within a given range around the central position of each fibre.
The disadvantages of using these spectra are that there is no cosmic-ray
rejection included in the extraction process and that the signal-to-noise
ratio drops rapidly as the spectra become fainter. 

 For fainter stars we tried using the {\sc giraffe} Base-Line Data Reduction
Software (girBLDRS) version 1.13.1 \citep{2000SPIE.4008..467B} to perform
optimal extraction of the spectra. This  maximizes the signal-to-noise in the
resulting spectra by weighting the pixels according to the variance of each
pixel and a model of the spatial profile \citep{1986PASP...98..609H}. Some
sort of weighted extraction  is essential to produce usable spectra for the
faintest stars in our dataset. It is also possible to identify pixels affected
by cosmic-rays in the images and exclude them from the extracted spectra by
comparing the spatial profile of the spectra with a model. Unfortunately, we
were not able to achieve the same level of stability in the radial velocities
for data extracted with girBLDRS as the normal pipeline extraction. We believe
that this is due to girBLDRS separately optimizing the profile used to extract
the object spectra, the flat-field spectra and the thorium-argon arc spectra.
The result is that the profiles used to extract the object, arc and flat
spectra are slightly different. This optimization is used to account for small
shifts in the positions of the spectra on the detector. The fringe-pattern can
change the effective detector efficiency by about 20\,percent over a spatial
scale of only a few pixels and it does not move relative to the detector.
Thus, using different profiles for the object, arc and flat spectra results in
inaccurate flat-fielding and the introduction of spurious high-frequency noise
into the spectra. This appears to be enough to reduce the precision of the
radial velocities that can be measured from these spectra to 2\,--\,3\,\kms.

 For these reasons we decided to develop our own method for extracting the
spectra. The key feature of the method is to use the flat-field images to
create an empirical model of the spatial profile for the spectra. This spatial
profile can then be used to perform a weighted extraction of the spectra and
to identify pixels affected by cosmic-rays in the images. The same weights can
be used for the object frames, flat-field frames and arc frames, so the
flat-fielding process does not introduce high-frequency noise into the
spectra, as is the case for optimal extraction with girBLDRS. One disadvantage
of this method is that small shifts in position between the flat-field images
and the object frames mean that the weights applied in the extraction are not
quite optimal. This results in a small reduction in the signal-to-noise ratio
of the extracted spectra.  A more important effect is that the flux at each
wavelength is slightly underestimated and that this flux-deficit varies with
wavelength and between spectra. For our {\sc giraffe} spectra we find that the
flux-deficit is a few percent and that it varies smoothly wavelength. This
only becomes a problem during the sky-subtraction phase of the data reduction.
The first step in sky subtraction is to form an average sky spectrum from
those fibres that were pointing at blank areas of the sky during the exposure.
It is normally possible to simply subtract this average sky spectrum from the
spectra of targets obtained in the same pointing. In our case, we first had to
calculate an optimum scaling factor for each spectrum to be applied to the
average sky spectrum prior to subtraction. This was done by finding the
scaling factor that minimized the root-mean square (RMS) difference between
the sky-subtracted spectrum and a smoothed version of this spectrum.

 We calculated our own dispersion solution for the spectra from the ThAr
spectra obtained on the same day as the actual observations. Typical shifts
over a 12 hour timescale for the {\sc giraffe} spectrograph are less than 0.2
pixels \citep{2004SPIE.5492..136P}. We used a 6-th order polynomial fit to the
positions of 22 unsaturated arc lines in the arc spectra. The worst RMS
residual  was 0.092\AA, the median RMS residual was 0.0045\AA. The mean
dispersion of the spectra is approximately 4.9\,\kms\,pixel$^{-1}$.

 In order to correct the spectra for telluric absorption we used synthetic
absorption spectra from a 6-layer model of the Earth's atmosphere
\citep{1988JQSRT..40..275N} and the {\sc hitran} molecular database
\citep{2005JQSRT..96..139R}. The parameters of the model were optimized by
fitting a {\sc uves} spectrum obtained at the same time as the {\sc giraffe}
observations. The optimum fit was achieved by minimizing the  mean absolute
deviation from a low-order polynomial fit to the {\sc uves} spectrum after
dividing through by the model spectrum. The fit to the telluric absorption was
very good (Fig.~\ref{TelluricFig}).  This synthetic telluric spectrum was
convolved with a Gaussian profile to match the resolution of the {\sc giraffe}
spectra. The synthetic telluric spectrum was then divided into the target
spectra. This removed all visible traces of telluric contamination. All {\sc
giraffe} spectra were interpolated onto a uniform velocity scale of
4.93\,\kms\,pixel$^{-1}$ using 3200 pixels covering the wavelength range
8061.1\,--\,8496.6\AA. We excluded spectra with a mean signal-to-noise ratio
less than 5 from our analysis and also excluded stars with fewer than two such
spectra. Examples of the resulting spectra for several targets are shown in
Fig.~\ref{ExampleSpectraFig}.

\section{Analysis}
 \label{Analysis}

 Stars are identified in this paper by the J2000 coordinates as listed in
\citet{2005MNRAS.356...89K} or \citet{2003A&A...404..171B} truncated to one
decimal place in right ascension and truncated to the nearest arcsecond in
declination.

\subsection{Radial velocity measurements}
 Radial velocities for all targets were measured using cross-correlation
against a template spectrum of the brown dwarf star UScoCTIO\,055. This star
is a visual binary with two similar components separated by only 0.12\,arcsec
and a combined spectral type of M5.5 \citep{2005ApJ...633..452K}. We  obtained
5 {\sc uves} spectra of this star from the ESO archive and formed the median
average spectrum. We re-binned this spectrum onto the same wavelength scale as
the {\sc giraffe} spectra and  convolved this spectrum with a Gaussian with
full-width at half-maximum (FWHM) of 3 pixels to match approximately the
resolution of these spectra.  Regions of the spectra affected by strong sky
line emission were excluded from the cross-correlation. The radial velocity
(RV) derived from the position of the peak of cross-correlation function (CCF)
and its error was measured using a parabolic fit to the three points at the
top of the CCF. The RV of UScoCTIO\,055 was taken to be $-6.38\kms$
\citep{2006MNRAS.372.1879K}.

\begin{table}
\begin{center}
 \caption{Radial velocities, \Vr, of stars in \sori\ and \lori.  The date
of observation is given as modified heliocentric Julian date (MHJD). The
standard error of the RV given here includes the ``external
error'' and zero-point correction measured from the sky lines described in the
text. The full-width at half-maximum of the cross-correlation function is
given in the final column. {\it The full version of this table is only
available in the on-line version of this paper.} \label{RVTable}}
\begin{tabular}{@{}rrrr}
\hline
  \multicolumn{1}{@{}l}{Star}  &
  \multicolumn{1}{@{}c}{MHJD}  &
  \multicolumn{1}{@{}l}{\Vr}  &
  \multicolumn{1}{@{}l}{FWHM}  \\
 &
  &
  \multicolumn{1}{@{}l}{(\kms)}  &
  \multicolumn{1}{@{}l}{(\kms)}  \\
\hline
J053557.0+094652 &53657.309  & 28.89  $\pm$ 0.67 &  63\\ 
J053557.0+094652 &53691.200  & 28.52  $\pm$ 0.60 &  63\\ 
J053557.0+094652 &53734.145  & 28.74  $\pm$ 0.68 &  63\\ 
J053539.4+095032 &53657.309  & 27.22  $\pm$ 0.68 &  69\\ 
J053539.4+095032 &53691.200  & 27.31  $\pm$ 0.61 &  70\\ 
J053539.4+095032 &53734.145  & 27.82  $\pm$ 0.69 &  68\\ 
J053530.4+095034 &53657.309  & 28.70  $\pm$ 0.71 & 116\\ 
J053530.4+095034 &53691.200  & 28.09  $\pm$ 0.66 & 116\\ 
J053530.4+095034 &53734.145  & 28.77  $\pm$ 0.72 & 113\\ 
\hline
\end{tabular}   
\end{center}    
\end{table}

\begin{table*}
\begin{center}
 \caption{Summary of measurements for each target.
{\it The full version of this table is only available in the on-line 
version of this paper.} \label{SummaryTable}}
\begin{tabular}{@{}rrrrrrr}
\hline
  \multicolumn{1}{@{}l}{Star}  &
  \multicolumn{1}{@{}c}{\Ic}  &
  \multicolumn{1}{@{}c}{$N_{\rm rv}$}  &
  \multicolumn{1}{@{}c}{\Vbar}  &
  \multicolumn{1}{@{}c}{$\log(p)$}  &
  \multicolumn{1}{@{}c}{EW(Na\,I)}&
  \multicolumn{1}{@{}c}{TiO(8442)} 
 \\

&
  \multicolumn{1}{c}{(mag)} &
 &
  \multicolumn{1}{c}{(\kms)} &
  & \multicolumn{1}{c}{(\AA)} 
\\
\hline
J053557.0+094652 &14.02 & 3 &$ 28.70 \pm 0.37$ &$-0.04 $&$ 2.34 \pm 0.01 $&$  0.677  \pm 0.001  $\\
J053539.4+095032 &14.06 & 3 &$ 27.44 \pm 0.38$ &$-0.10 $&$ 2.34 \pm 0.02 $&$  0.667  \pm 0.002  $\\
J053530.4+095034 &14.10 & 7 &$ 28.98 \pm 0.44$ &$-1.76 $&$ 2.33 \pm 0.01 $&$  0.668  \pm 0.001  $\\
J053502.7+095647 &14.16 & 4 &$ 30.06 \pm 1.01$ &$-4.57 $&$ 2.33 \pm 0.02 $&$  0.663  \pm 0.001  $\\
J053408.4+095125 &14.17 & 4 &$  6.52 \pm 0.34$ &$-0.05 $&$ 2.61 \pm 0.02 $&$  0.637  \pm 0.001  $\\
J053426.0+095149 &14.36 & 4 &$ -0.65 \pm 0.34$ &$-0.08 $&$ 2.85 \pm 0.01 $&$  0.657  \pm 0.001  $\\
J053555.6+095053 &14.38 & 3 &$ 27.26 \pm 0.39$ &$-0.08 $&$ 2.53 \pm 0.02 $&$  0.682  \pm 0.002  $\\
\hline
\end{tabular}   
\end{center}    
\end{table*}

 The precision with which the peak of the CCF can be measured is unlikely to
be a true reflection of the accuracy with which we can measure the radial
velocities from our spectra. In order to quantify the accuracy of our radial
velocities we  measured the radial velocities of the night sky emission lines
in our spectra. We used the night sky emission line spectrum observed with the
{\sc uves} spectrograph as a template in the cross-correlation
\citep{2003A&A...407.1157H}. We measured all the spectra observed at a given
pointing, including object spectra prior to sky-subtraction, and calculated
the mean and standard deviation of the resulting radial velocities. The
standard deviation, $\sigma_{\rm sky}$, is given in Table~\ref{ObsDateTable}.
The mean RV shift of the sky-line spectra for a given pointing
ranges from $-1.28$\,\kms\ to $-0.25$\,\kms. We have subtracted this mean radial
velocity shift from the measured radial velocities of our targets, i.e., the
night sky emission line spectrum is used to define the zero-point of the
stellar radial velocities at each pointing. We have also added the
value of $\sigma_{\rm sky}$ for each frame in quadrature to the precision of
the stellar RV measured from the cross correlation
function. For 85\,percent of our RV measurements $\sigma_{\rm
sky}$ is the dominant term in the uncertainty. The resulting RV
measurements with their standard deviations are given in Table~\ref{RVTable}. 

\begin{figure}
\includegraphics[width=0.47\textwidth]{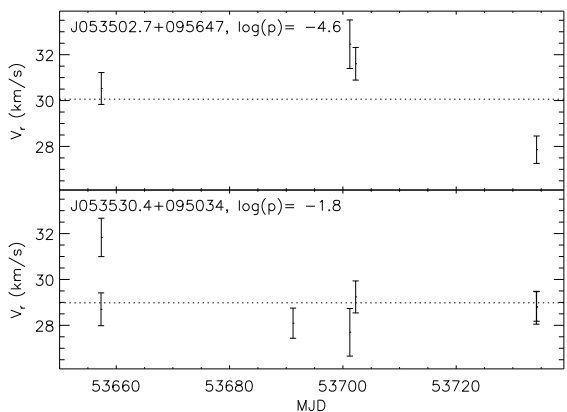}
\caption{\label{RVPlotFig}Example radial velocity measurements as a function
of time. The star identifier and value of $\log(p)$ are indicated in each
panel. The weighted mean radial velocity is indicated by a dotted line.
}
\end{figure}

\begin{figure}
\includegraphics[width=0.47\textwidth]{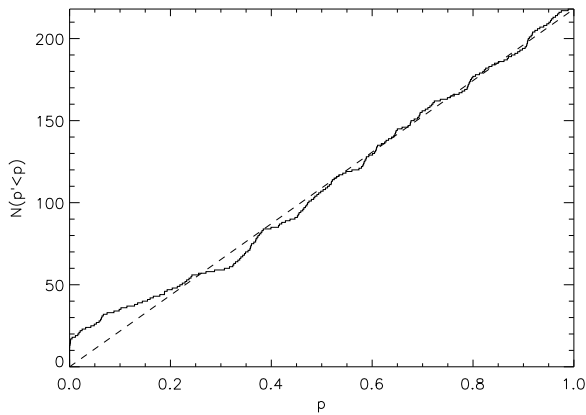}
\caption{\label{logp_cdf}Left panel: Cumulative distribution of $p$. 
The difference between this distribution and a uniform distribution (dashed
line) is not significant. }
\end{figure}
\begin{figure}
\includegraphics[width=0.47\textwidth]{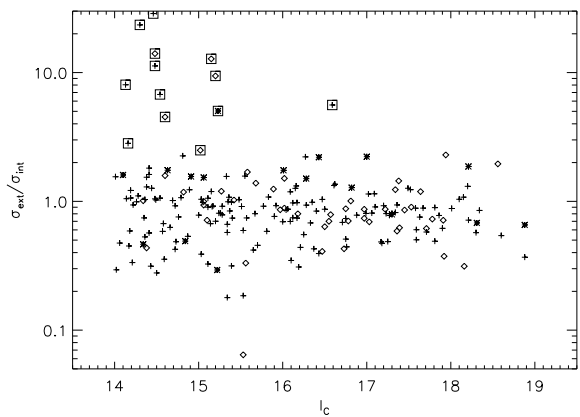}
\caption{\label{ZFig}
 The ratio of the standard errors of the mean radial velocities, $\sigma_{\rm
ext}$ and $\sigma_{\rm int}$, as a function of \Ic\  magnitude. The width of
the CCFs is indicated as follows: crosses, FWHM $<$ 80\,\kms; diamonds,
80\,\kms $<$ FWHM $<$ 100\,\kms; asterisks, FWHM $>$ 100\,\kms.  Stars with
significantly variable radial velocities are highlighted (squares). }
\end{figure}

\subsection{Variability criterion for radial velocities}
 For each star we have $N_{\rm rv}$ RV measurements $V_{{\rm
r},i}$ each with standard error $\sigma_i$. We calculate the weighted mean
RV for each star, $\Vbar$, and then calculate the chi-squared
statistic for $\Vbar$ as a model for the observed radial velocities, i.e., 
\[ \chi^2 = \sum^{N_{\rm rv}}_{i=1} \frac{(V_{{\rm r},i} - \Vbar)^2}
{\sigma_i^2} \]
 In order to identify stars with variable radial velocities  we calculate the
probability $p$ of observing this value of $\chi^2$ or greater from a sample
of normally distributed random observations with mean  \Vbar\ and standard
errors $\sigma_i$. Our criterion for identifying stars with variable radial
velocities is $\log(p)< -4$. The probability that one or more stars are
incorrectly identified as having variable radial velocities by chance due to
statistical fluctuations (assumed to be normally distributed) in our sample of
\Ntarget\ stars is about 2\,percent. The values of $N_{\rm rv}$, $\log(p)$ and
\Vbar\ for each target are given in Table~\ref{SummaryTable}. There are two
ways to calculate the standard error of the weighted mean, the external error
based on the scatter of the data,
\[\sigma_{\rm ext} = \sqrt{\frac{\chi^2}{(N_{\rm rv}-1)\sum 1/\sigma_i^2}},\]
and the internal error based on the standard errors only,
\[\sigma_{\rm int} = \sqrt{\frac{1}{\sum 1/\sigma_i^2}}.\]
The value given in  Table~\ref{SummaryTable} is the larger of these two
values.

 There are \Nvariable\ stars in our sample that have variable radial
velocities according to our criterion $\log(p) < -4$. The properties of these
stars are listed in Table~\ref{BinaryTable}. Stars in which  both components
are visible in the spectra or CCFs are identified as SB2, those showing only a
single spectrum are identified as SB1. Examples of radial velocities for stars
with $\log(p)< -4$ and $\log(p)> -4$ are shown in Fig.~\ref{RVPlotFig}. 

 If our values of $\sigma_i$ are good estimates of the true uncertainty on
each RV measurement and the majority of stars in our sample are non-variable,
then we would expect that the distribution of  $p$ will be uniform. The
cumulative distribution of our measured $p$ values is compared to a
uniform distribution in Fig.~\ref{logp_cdf}. We have tested the hypothesis
that these two distribution are equal using the Kolmogorov-Smirnov test and
find that there is no evidence for any significant difference between them.
 
 We have also tested the reliability of our  $\sigma_i$ values by considering
the ratio $Z=\sigma_{\rm ext}/\sigma_{\rm int}$. If the values of $\sigma_i$
are reliable then we expect $Z\approx 1$. More precisely, for samples drawn
from normal distributions the mean value of Z is 1 with standard error
$1/\sqrt{2(N_{\rm rv}-1)}$ \citep{Topping}. The  values of $Z$ for our RV
measurements are shown as a function of \Ic\ magnitude in Fig.~\ref{ZFig}. It
can be seen that the values of $Z$ are indeed close to 1 (with the exception
of the spectroscopic binaries, of course) and that there is no significant
trend of $Z$ with \Ic. Stars with broad spectral lines are highlighted in this
figure so that it can also be seen that the values of $Z$ are not
significantly different for rapidly rotating stars compared to other stars in
the sample.

\subsection{Binary stars identified from variable line width\label{fwhmsec}}
\begin{figure}
\includegraphics[width=0.47\textwidth]{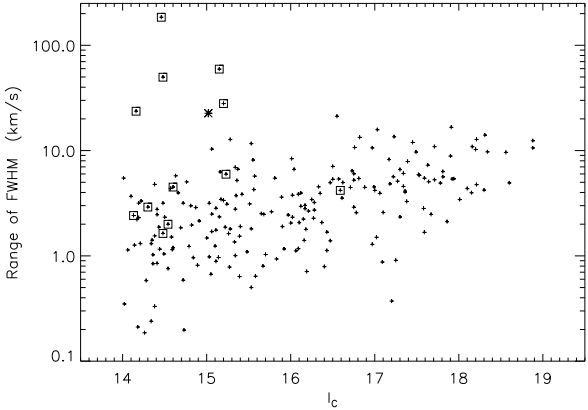}
\caption{\label{FWHMFig}The range in full-width at half maximum (FWHM) of the
cross-correlation function (CCF) for each target as a
function of \Ic\ magnitude. Stars with variable radial velocities are
highlighted (squares). The SB2 binary J054001.0$-$021959  discussed in the
text is marked with an asterisk.}
\end{figure}

 In Fig.~\ref{FWHMFig} we show the range in the FWHM of the CCFs against \Ic\
magnitude. We have used this plot to identify potential SB2 binary stars in
which blending of the spectra from two similar components results in
variations in the width of the spectral lines with little change in the radial
velocity measured from the peak of the CCF. Several of the SB2 binaries
identified above are recovered by this method. The star J054001.0$-$021959 has
a large range in FWHM compared to other stars of the same \Ic\ magnitude. The
four spectra of this star all have high signal-to-noise. There is a definite
broadening and asymmetry in one spectrum (Fig.~\ref{SpecFig}). In addition,
the value of $\log(p)$ for this star is close to our criterion for variable
radial velocities. This star appears to be an SB2 binary. One other star
(J053522.5+094501) has a range in FWHM $> 20$\,\kms, but this is a result of
two spectra of the seven obtained having low signal-to-noise. There is no sign
of asymmetry in the CCF and the radial velocities of this star show no hint of
variation ($\log(p) = -0.09$). Despite the large range of FWHM measured for
this star, there is no strong evidence that it an SB2 binary star.

\begin{figure*}
\begin{center}
\mbox{\includegraphics[angle=270,width=0.41\textwidth]{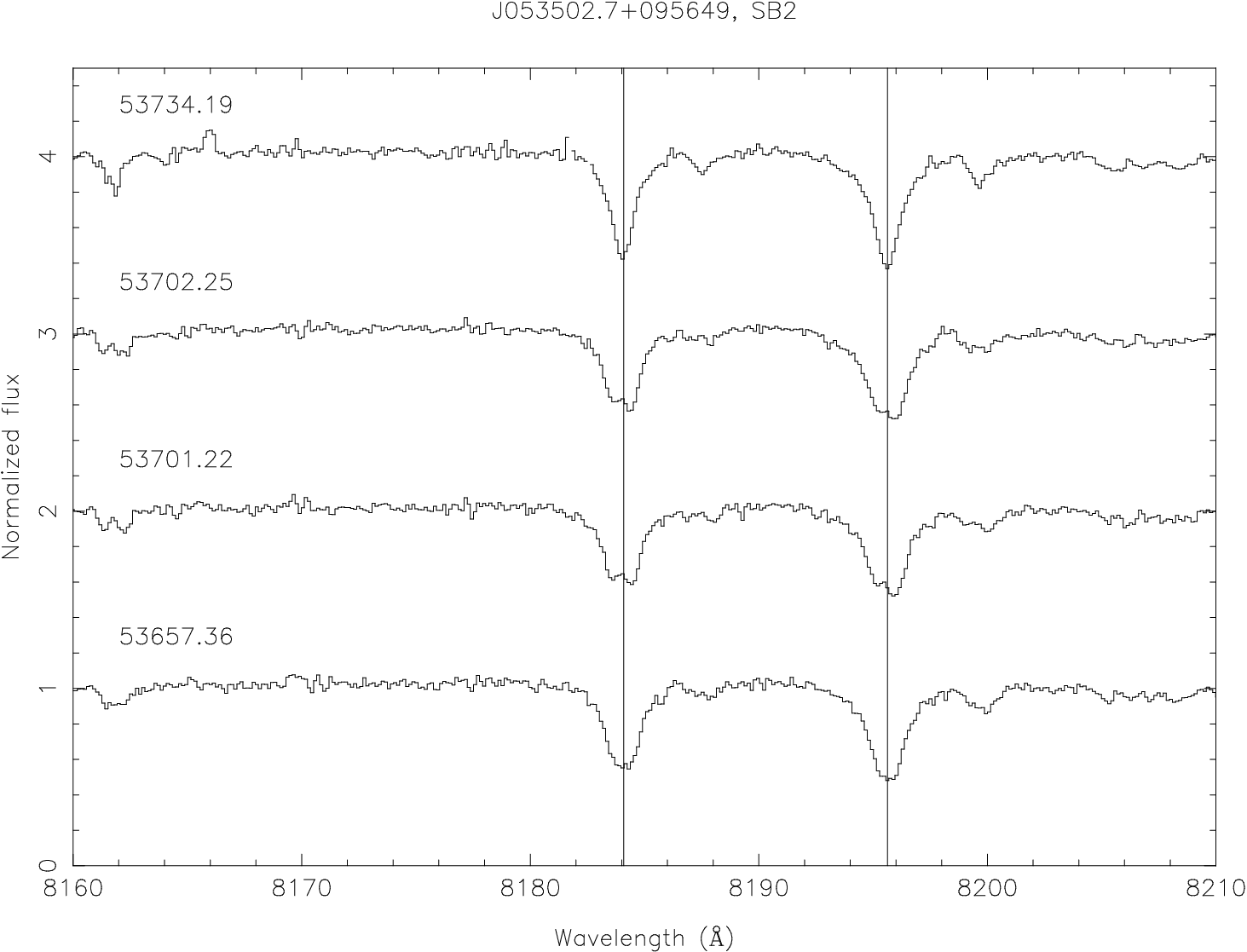}
      \includegraphics[angle=270,width=0.41\textwidth]{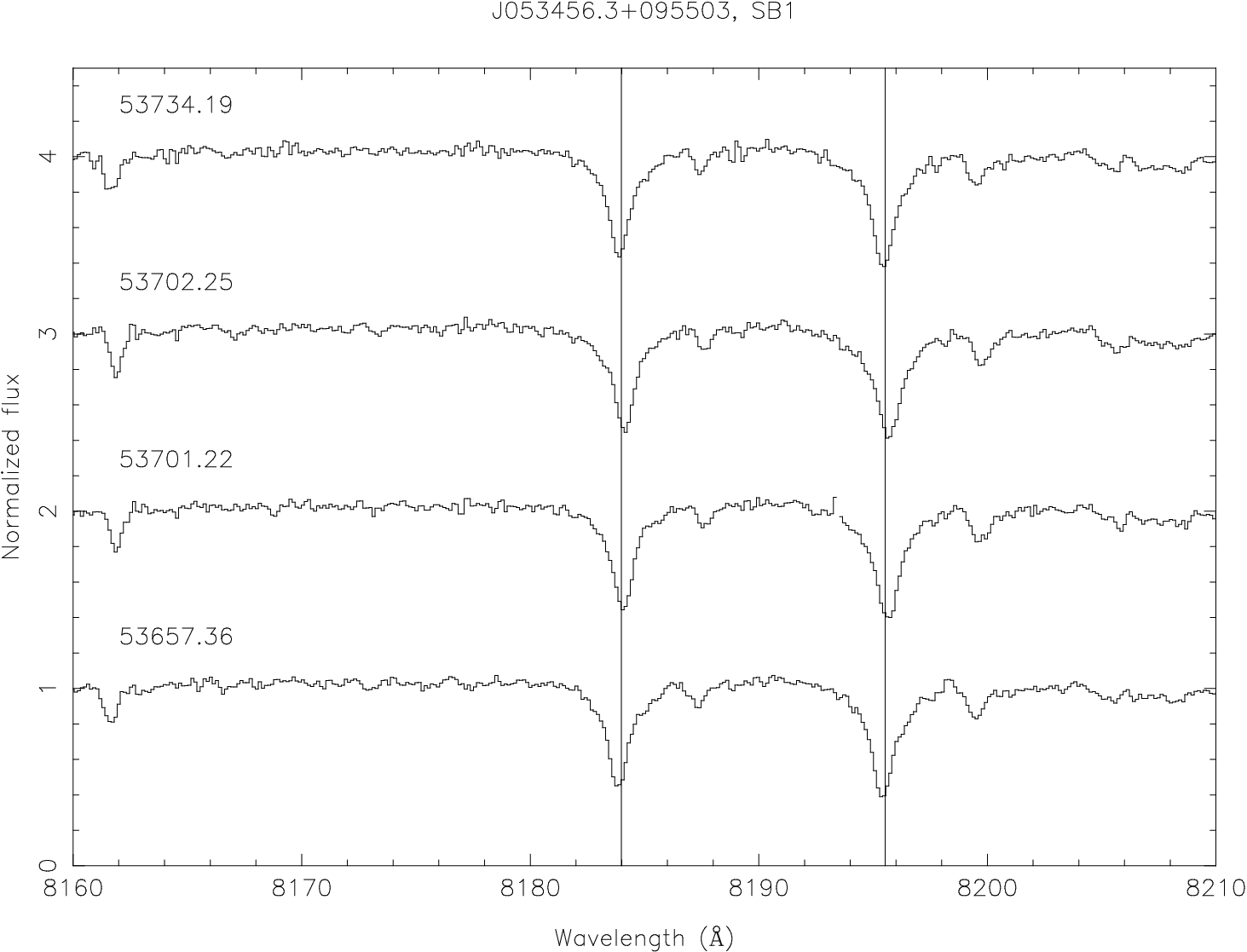}}
\mbox{\includegraphics[angle=270,width=0.41\textwidth]{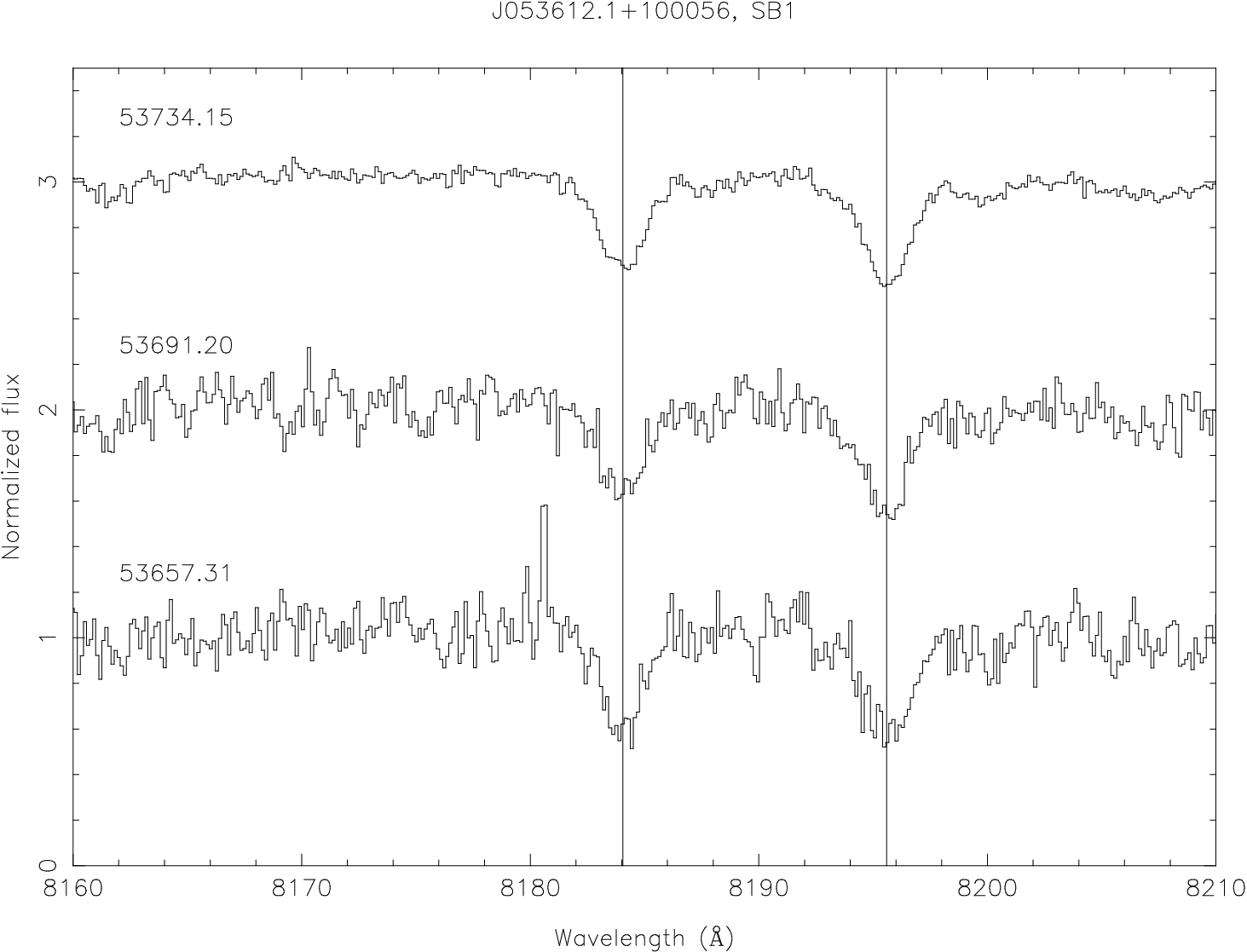}
      \includegraphics[angle=270,width=0.41\textwidth]{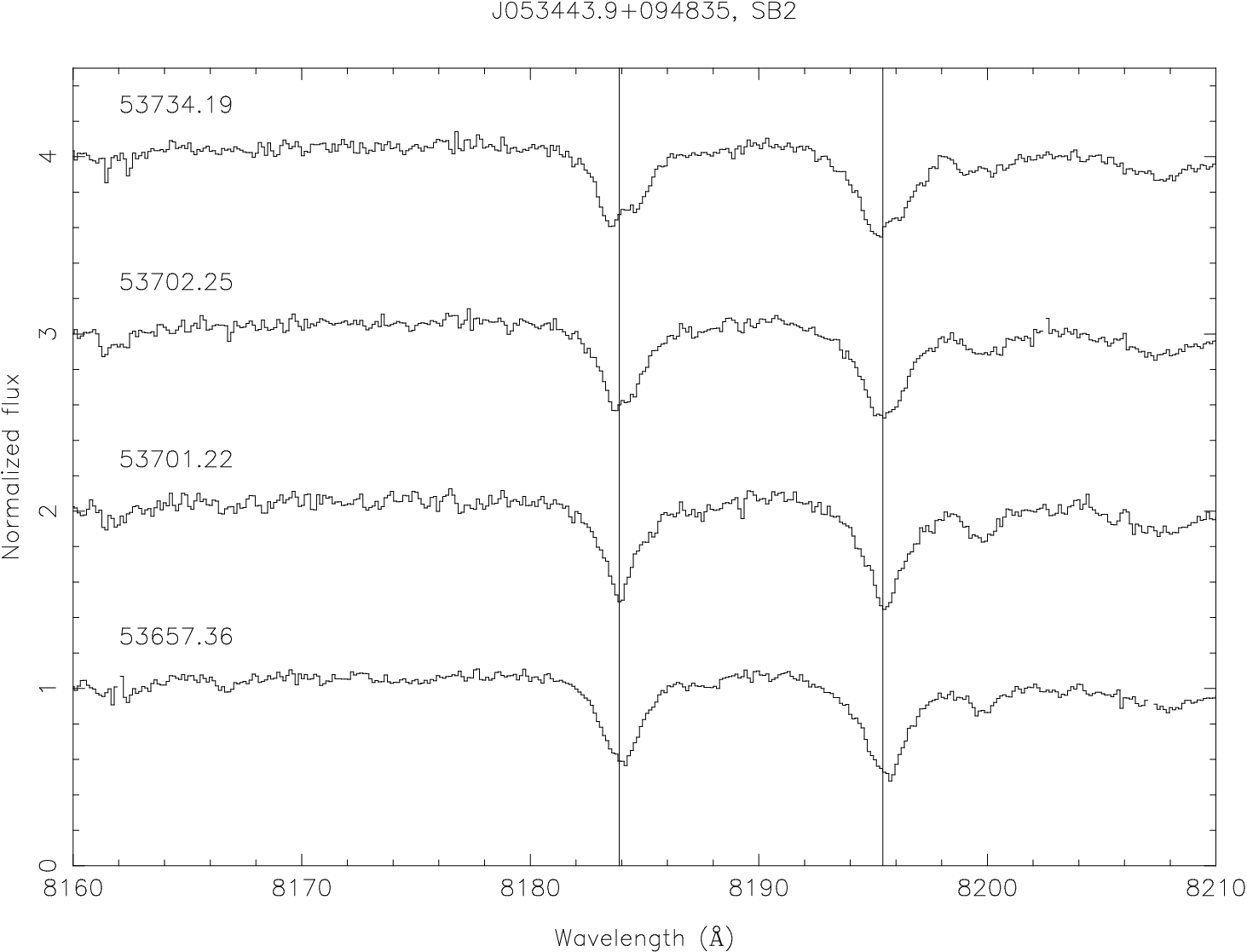}}
\mbox{\includegraphics[angle=270,width=0.41\textwidth]{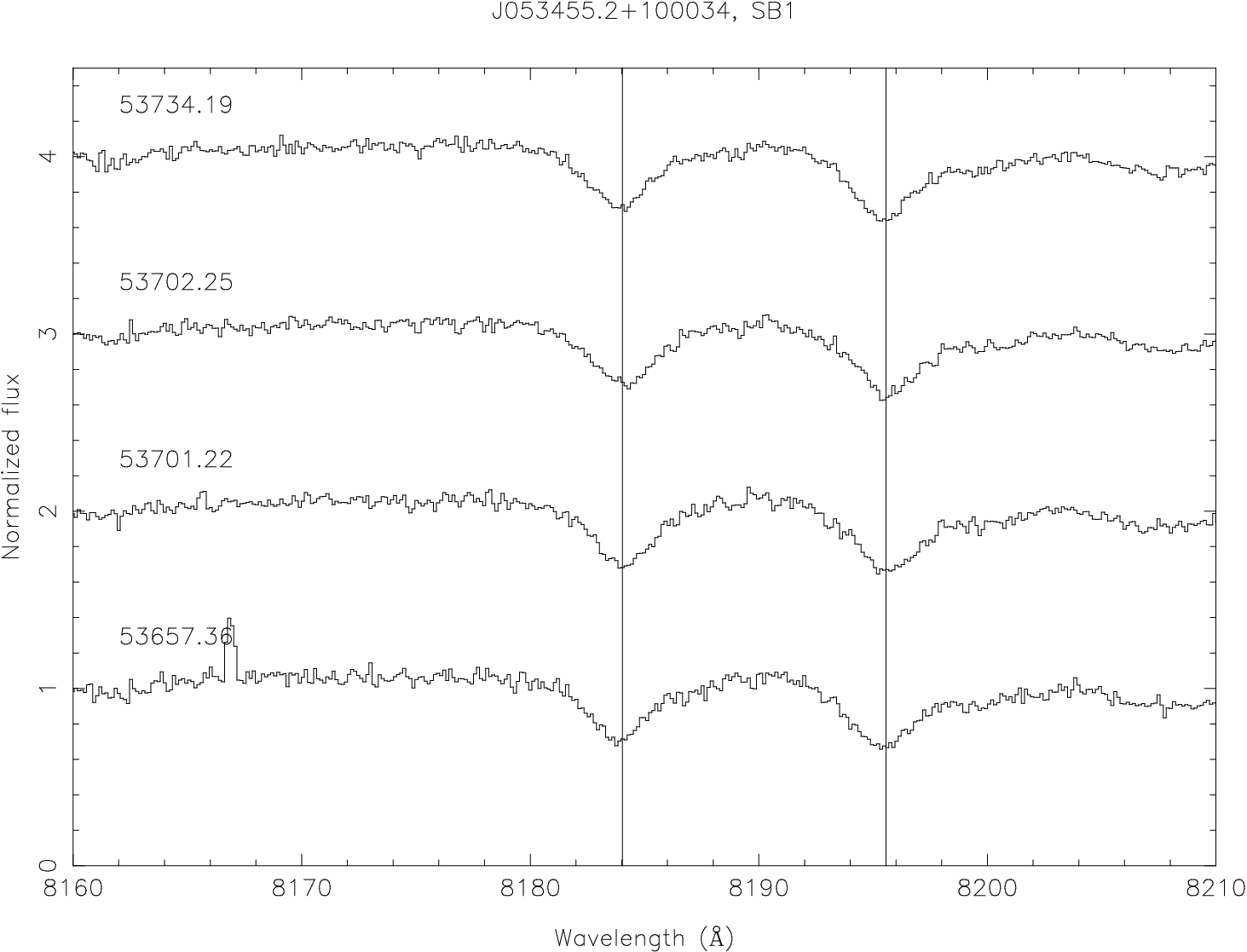}
      \includegraphics[angle=270,width=0.41\textwidth]{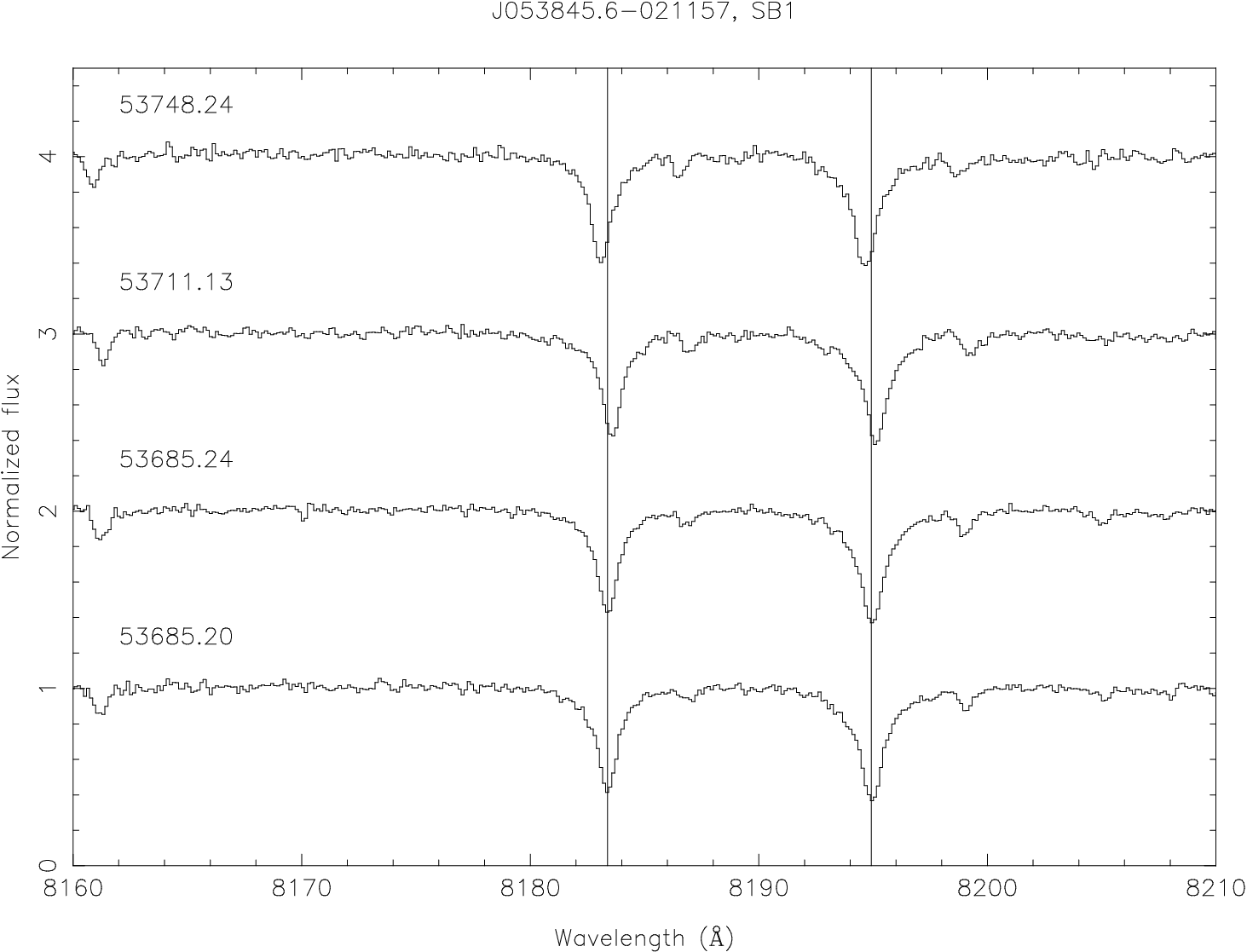}}
\mbox{\includegraphics[angle=270,width=0.41\textwidth]{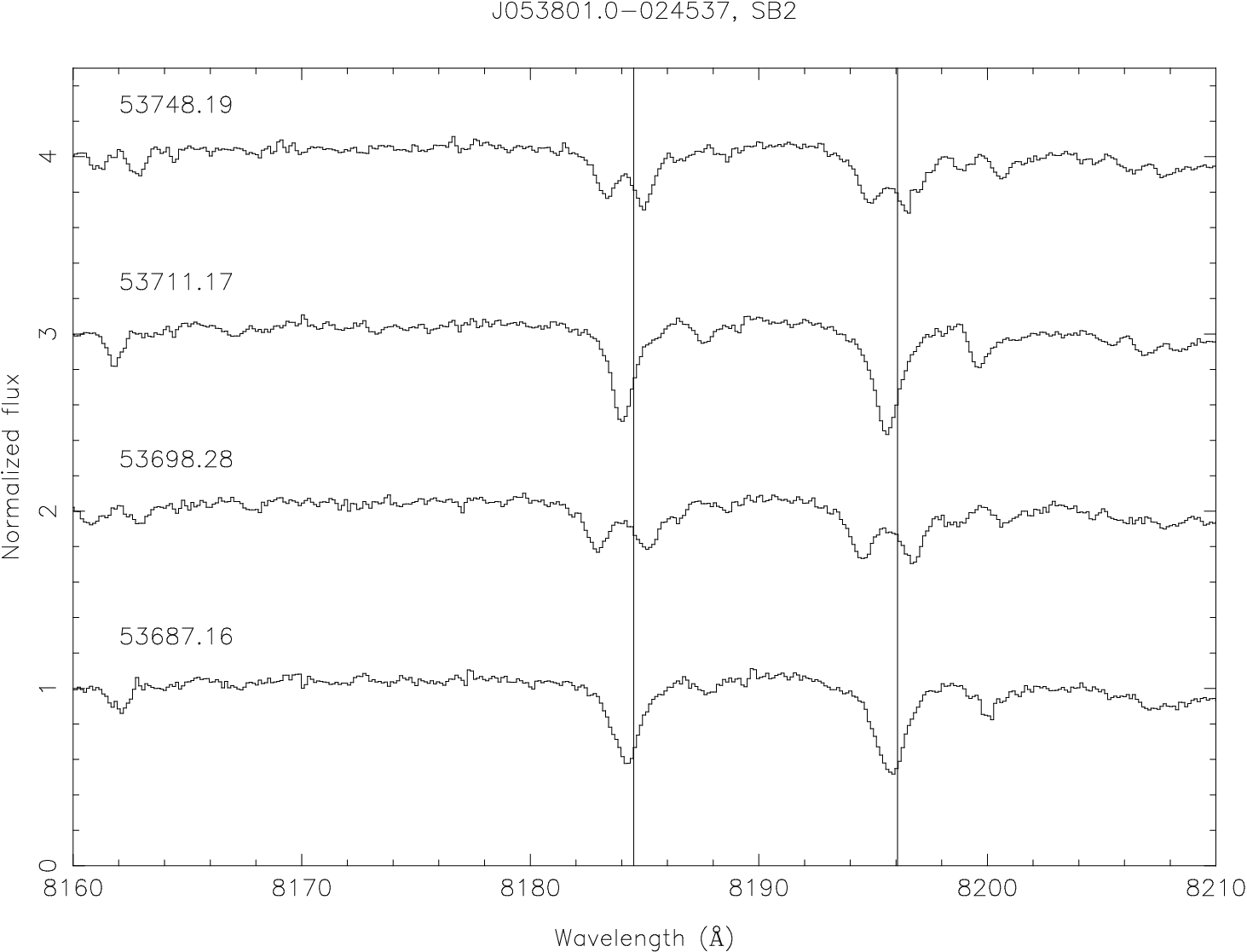}
      \includegraphics[angle=270,width=0.41\textwidth]{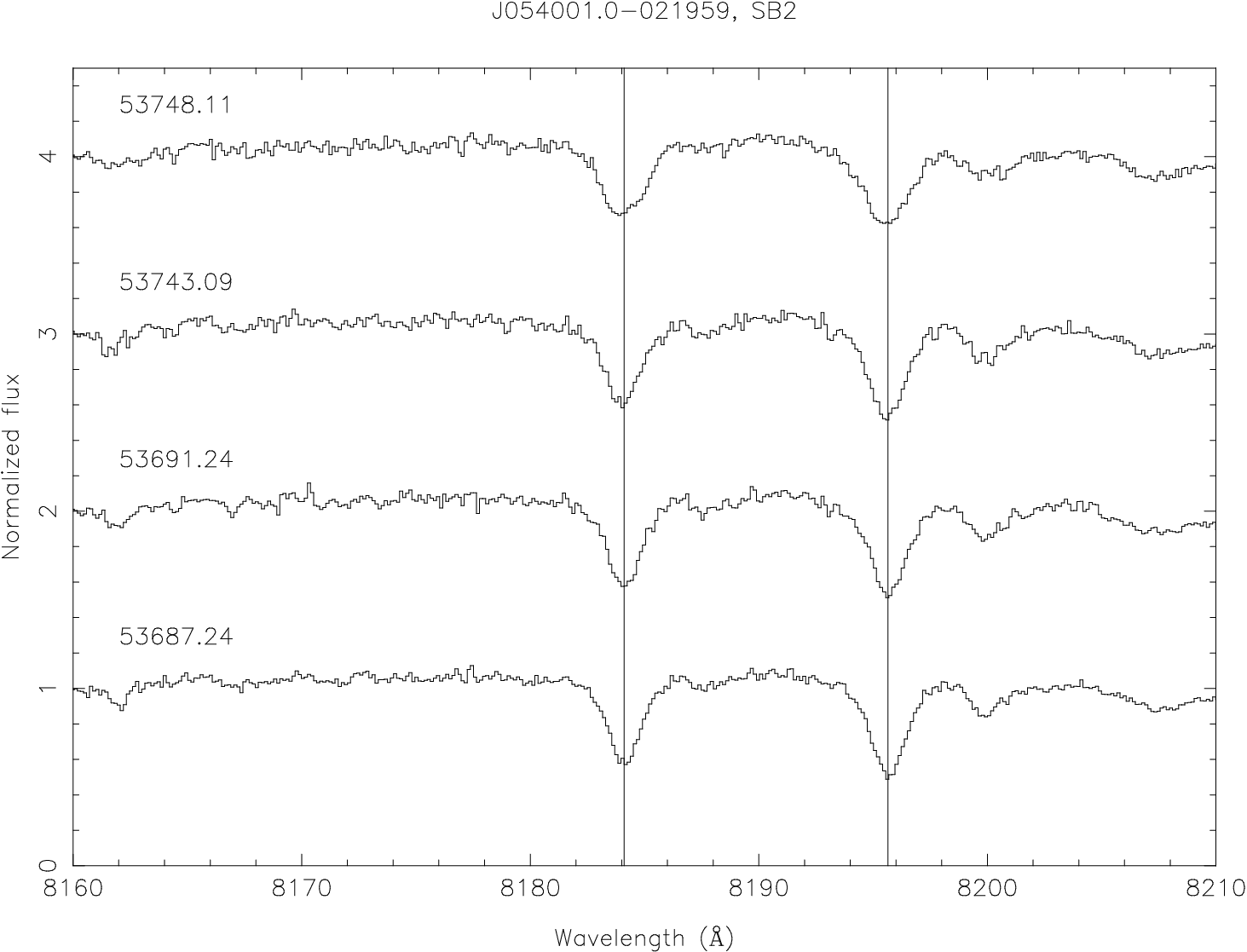}}
\end{center}                                          
\caption{\label{SpecFig} Spectra of the spectroscopic binaries discovered by
our survey around the Na\,I~8200 feature. The spectra have been normalized
using a low-order polynomial and offset for clarity. Spectra are labeled by
the modified heliocentric JD of observation. The wavelengths of the Na\,I doublet 
at the mean radial velocity of the star are marked by  vertical lines.}  
\end{figure*}

\begin{figure*}
\begin{center}
\mbox{\includegraphics[angle=270,width=0.41\textwidth]{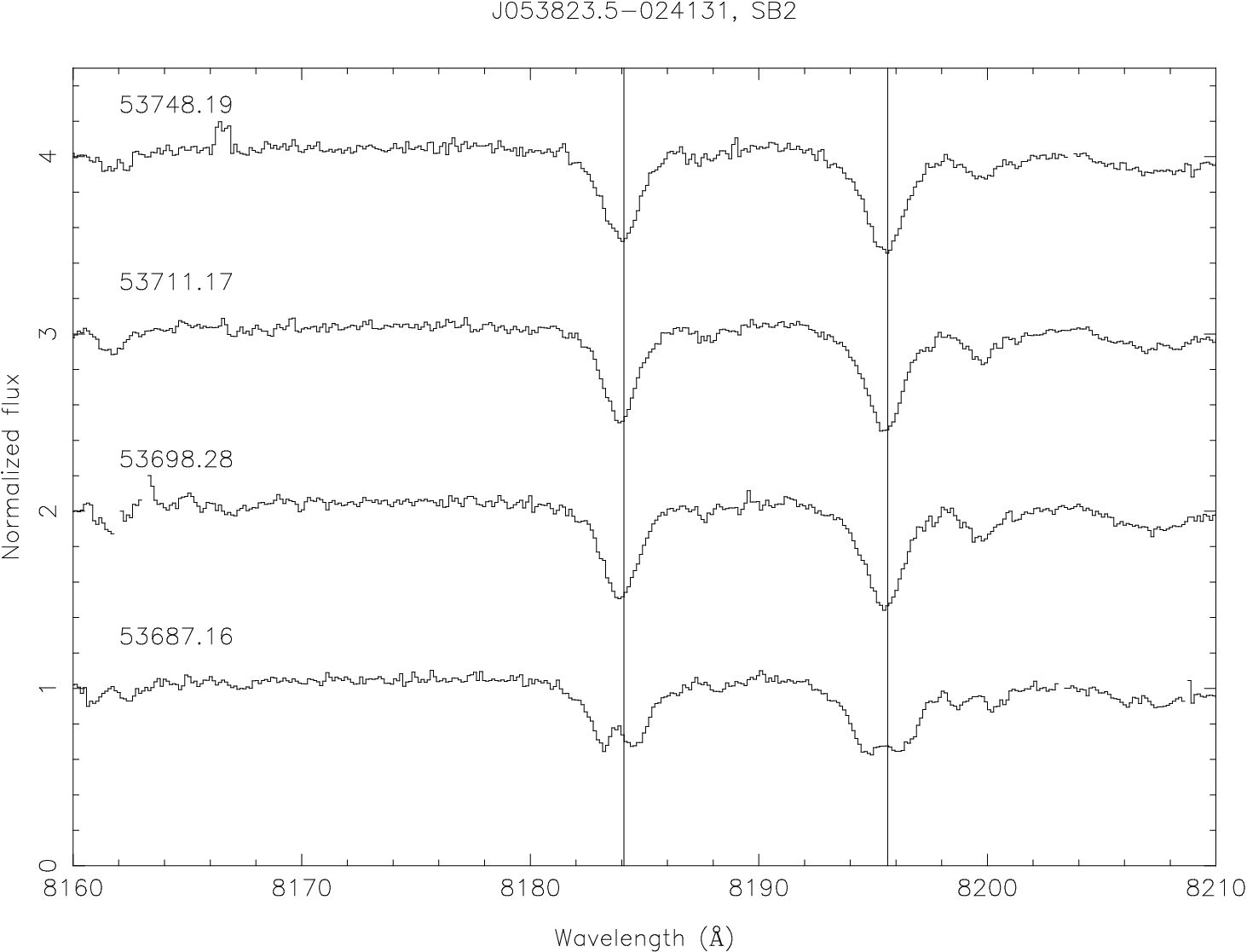}
      \includegraphics[angle=270,width=0.41\textwidth]{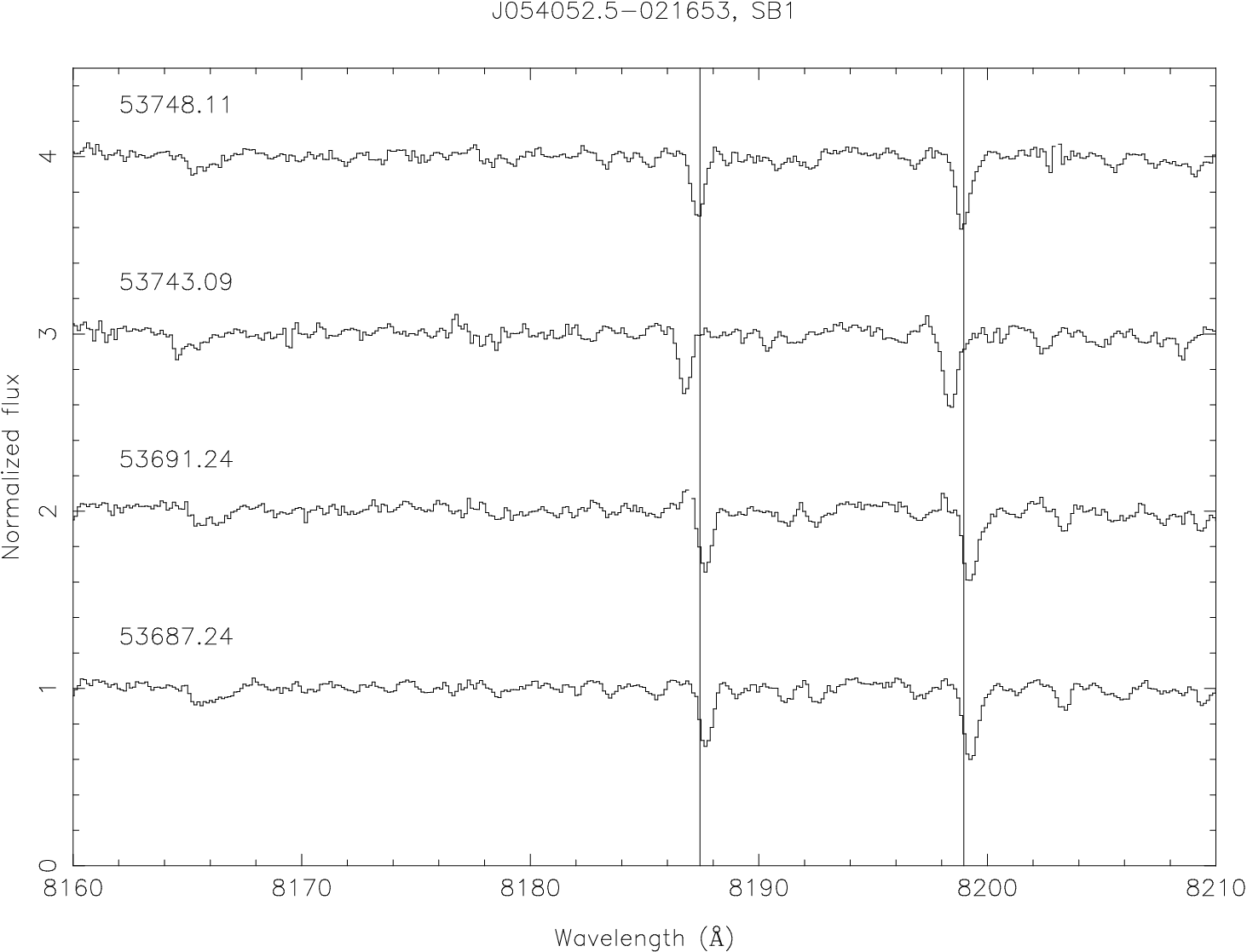}}
\mbox{\includegraphics[angle=270,width=0.41\textwidth]{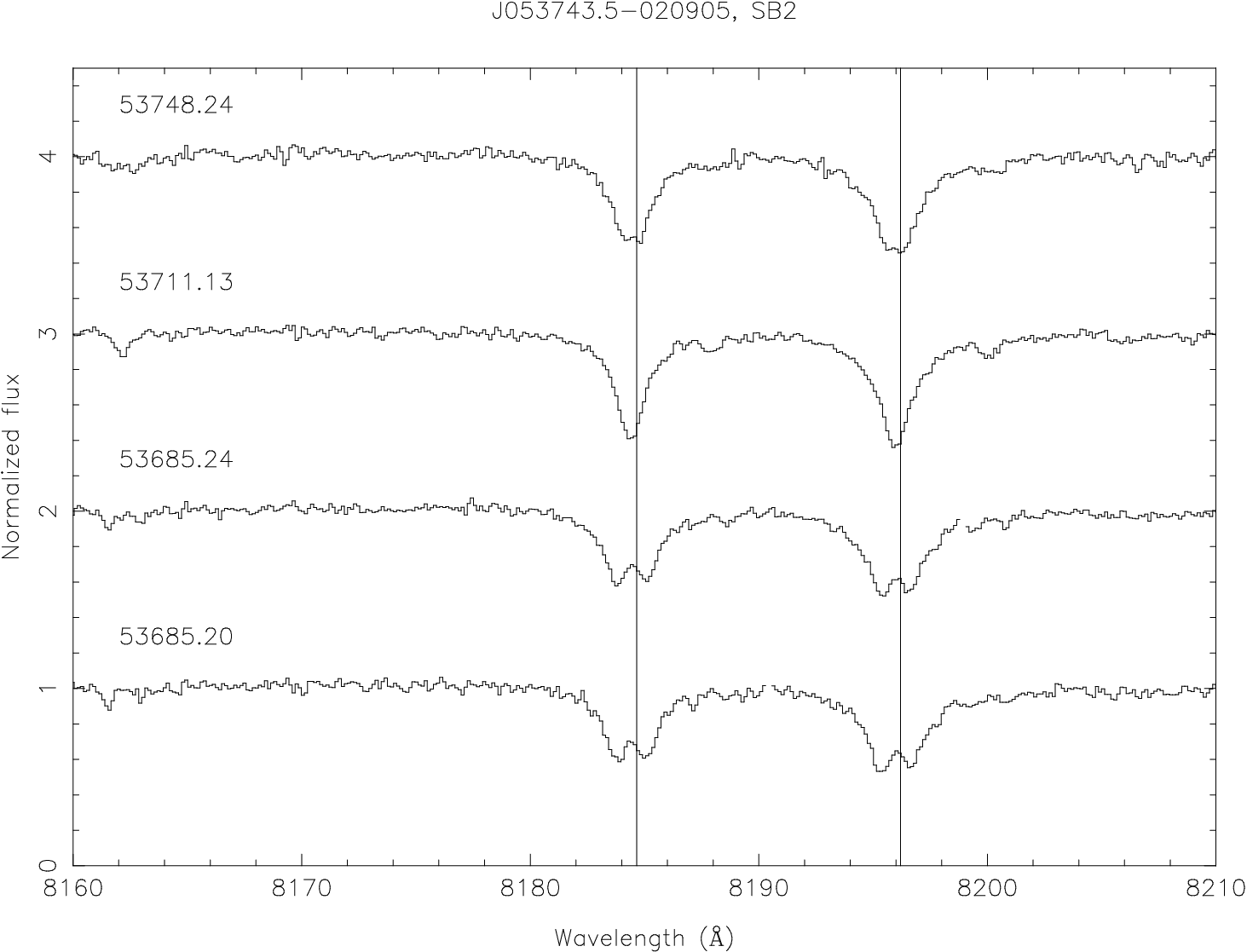}
      \includegraphics[angle=270,width=0.41\textwidth]{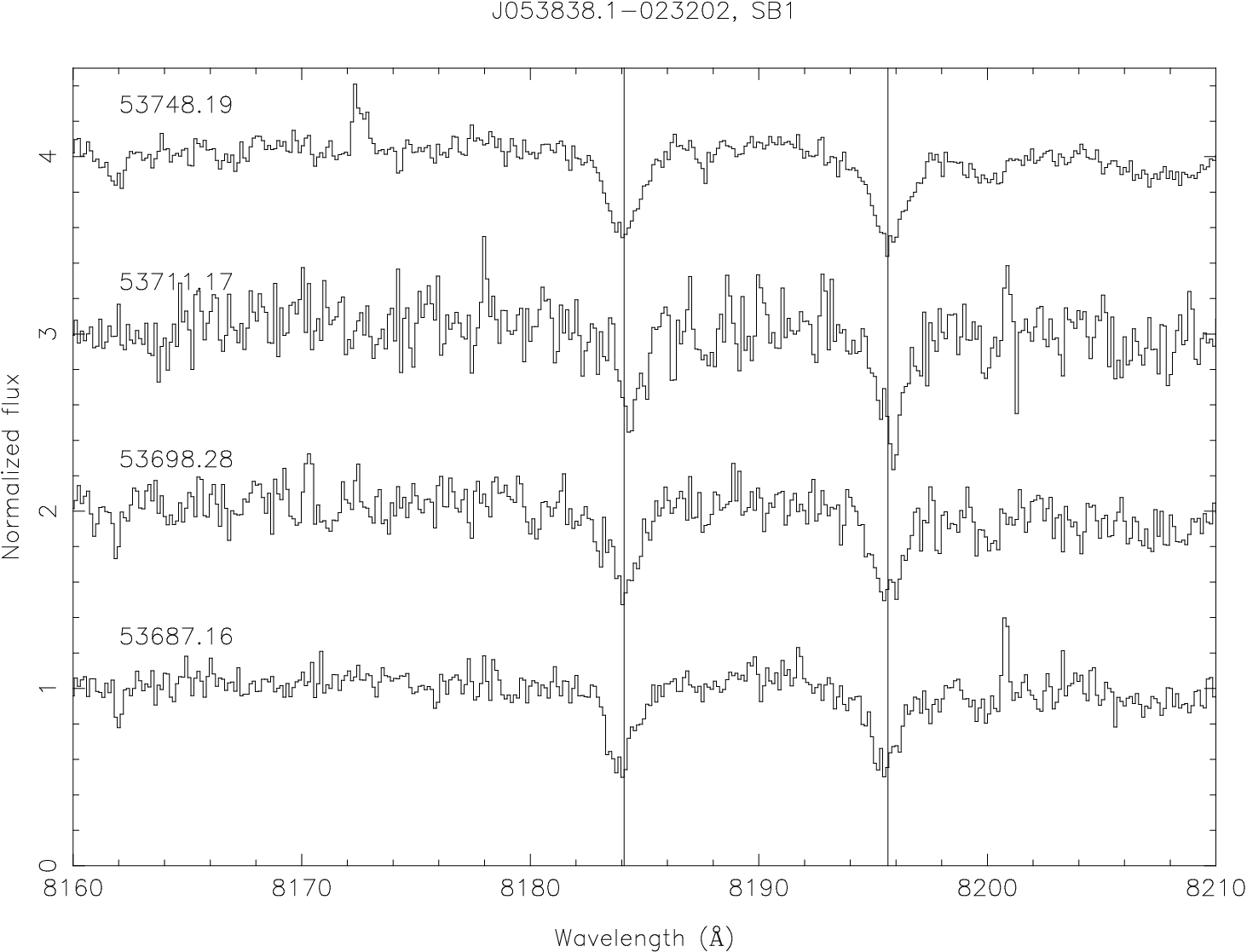}}
\mbox{\includegraphics[angle=270,width=0.41\textwidth]{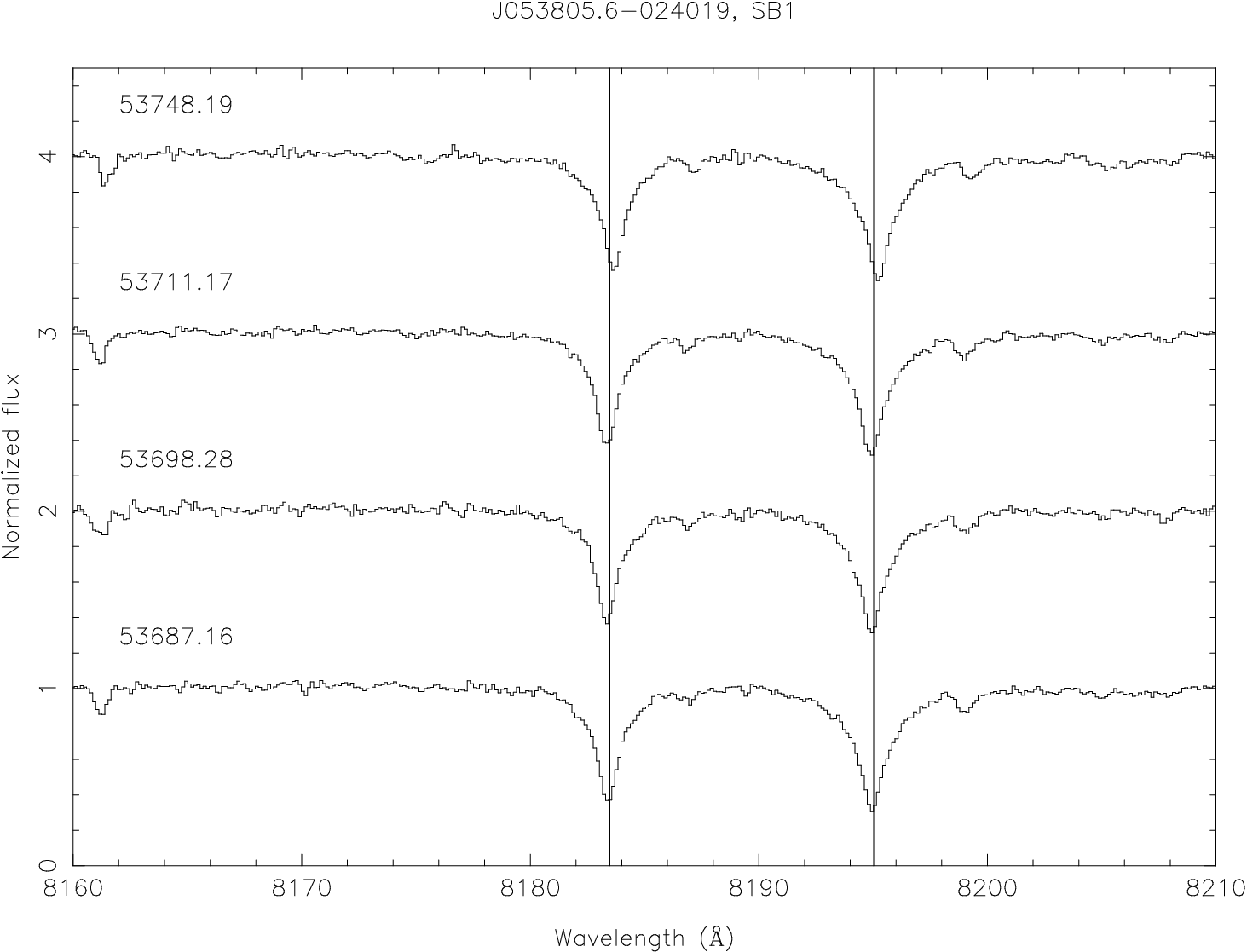}
      \includegraphics[angle=270,width=0.41\textwidth]{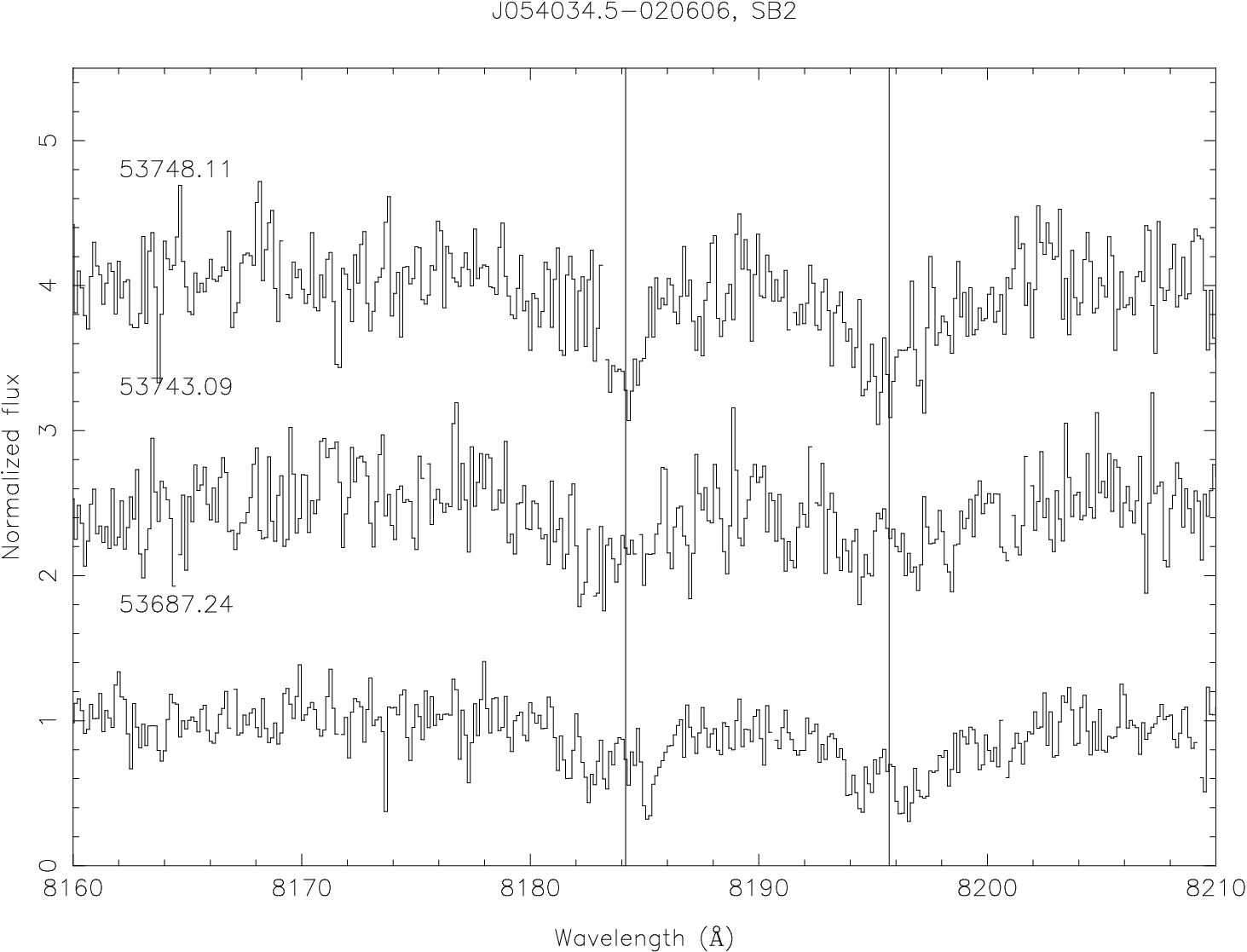}}
\end{center}
\contcaption{~}
\end{figure*}
 
\begin{table*} \begin{center} \begin{minipage}{165mm} \caption{Properties of
the binaries discovered by our survey. The standard error in the mean for
the measured radial velocities is listed under $\sigma_{rv}$. Non-members of
the clusters are indicated in the final column by the note `nm'.
\label{BinaryTable}}
\begin{tabular}{@{}llrrrrrrrl} \hline \multicolumn{1}{@{}l}{Star}  &
\multicolumn{1}{@{}c}{I$_{\rm C}$}  & \multicolumn{1}{@{}c}{$\Rc-\Ic$}  &
\multicolumn{1}{@{}c}{$N_{rv}$} & \multicolumn{1}{@{}c}{\Vbar } &
\multicolumn{1}{@{}c}{$\sigma_{rv}$} & \multicolumn{1}{@{}c}{$\log(p)$} &
\multicolumn{1}{@{}c}{EW(Na\,I)}& \multicolumn{1}{@{}c}{TiO(8442)} &
\multicolumn{1}{@{}l}{Notes} \\

  \multicolumn{1}{@{}c}{}  &
  \multicolumn{1}{@{}c}{(mag)}  &
  \multicolumn{1}{@{}c}{(mag)}  &
  \multicolumn{1}{@{}c}{} &
  \multicolumn{1}{@{}c}{(\kms)} &
  \multicolumn{1}{@{}c}{(\kms)} &
  \multicolumn{1}{@{}c}{} &
  \multicolumn{1}{@{}c}{(\AA)}&
  \multicolumn{1}{@{}c}{} &
  \multicolumn{1}{@{}c}{} \\
\hline
J053502.7$+$095649 & 14.16 & 1.30 & 4 & 30.06 &  1.01 &  $-4.6$& 2.33 $\pm$ 0.02 & 0.663 $\pm$ 0.001 &SB2 \\ 
J053456.3$+$095503 & 14.54 & 1.36 & 4 & 26.20 &  2.32 & $-28.8$& 2.21 $\pm$ 0.01 & 0.656 $\pm$ 0.001 &SB1 \\ 
J053612.1$+$100056 & 14.60 & 1.31 & 3 & 28.66 &  2.68 &  $-8.9$& 2.23 $\pm$ 0.10 & 0.666 $\pm$ 0.008 &SB1 \\ 
J053443.9$+$094835 & 15.20 & 1.69 & 4 & 22.56 &  3.48 &  $<-42$& 2.66 $\pm$ 0.02 & 0.732 $\pm$ 0.002 &SB2 \\ 
J053455.2$+$100034 & 15.23 & 1.72 & 4 & 27.92 &  2.20 & $-15.6$& 2.37 $\pm$ 0.02 & 0.710 $\pm$ 0.002 &SB1
\footnote{Hint of companion in CCF.} \\ 
J053845.6$-$021157 & 14.48 & 1.18 & 4 &  4.11 &  3.52 &  $<-42$& 1.99 $\pm$ 0.02 & 0.633 $\pm$ 0.001 &SB1 \\ 
J053801.0$-$024537 & 14.46 & 1.57 & 4 & 46.12 &  9.20 &  $<-42$& 2.09 $\pm$ 0.01 & 0.653 $\pm$ 0.001 &SB2 \\ 
J054001.0$-$021959 & 15.02 & 1.90 & 4 & 30.80 &  0.90 & $-3.5 $& 2.38 $\pm$ 0.02 & 0.726 $\pm$ 0.003 &SB2 
\footnote{Identified from variation in width of CCF.} \\ 
J053823.5$-$024131 & 15.15 & 1.69 & 4 & 30.23 &  4.26 &  $<-42$& 2.71 $\pm$ 0.02 & 0.690 $\pm$ 0.002 &SB2 \\ 
J054052.5$-$021653 & 14.30 & 1.13 & 4 &152.28 &  8.25 &  $<-42$& 0.52 $\pm$ 0.02 & 0.628 $\pm$ 0.001 &SB1, nm \\ 
J053743.5$-$020905 & 14.48 & 1.36 & 4 & 51.02 &  4.76 &  $<-42$& 2.65 $\pm$ 0.02 & 0.630 $\pm$ 0.001 &SB2 \\ 
J053838.1$-$023202 & 16.59 & 1.73 & 4 & 30.79 &  2.65 & $-19.5$& 2.05 $\pm$ 0.07 & 0.659 $\pm$ 0.006 &SB1 \\ 
J053805.6$-$024019 & 14.13 & 1.21 & 4 &  7.68 &  2.53 & $-41.1$& 2.43 $\pm$ 0.01 & 0.644 $\pm$ 0.001 &SB1 \\ 
\hline
\end{tabular}   
\end{minipage}
\end{center}    
\end{table*}     

\subsection{Membership criteria}
\label{membership}
\label{RVSelection}

\begin{figure}
\includegraphics[width=0.47\textwidth]{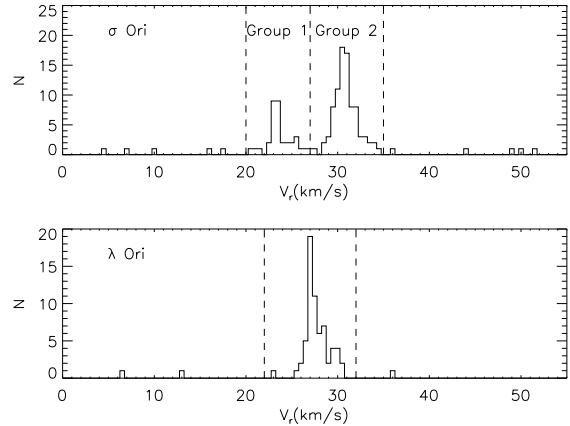}
\caption{\label{RVDistFig}The distributions of weighted mean radial velocity
for our targets in \lori\ and \sori. Dashed lines shows the selection criteria
for assigning stars to Group 1 or Group 2 for  \sori\ and for selecting
non-members for both clusters, as described in section \ref{RVSelection}.}
\end{figure}
\begin{figure}
\includegraphics[width=0.47\textwidth]{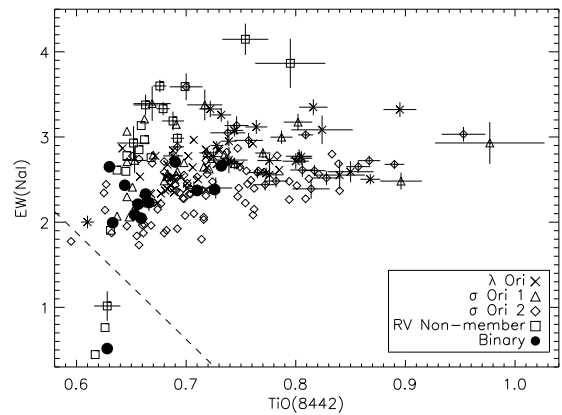}
\caption{\label{EWTiOFig} The value of EW(Na\,I) versus the TiO(8442) spectral
index for our targets. The plotting symbols used indicate binarity or
membership of the \sori\ or \lori\ clusters  or non-membership of these
clusters based on the radial velocity (if non-variable), as indicated in the
legend.  Stars below the dashed line are considered to be non-members. Error
bars are only shown in cases where they are larger than the plotting symbol
used.}
\end{figure}
\begin{figure}
\includegraphics[width=0.47\textwidth]{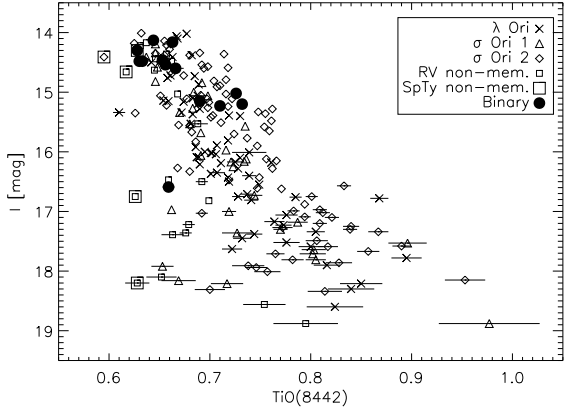}
\caption{\label{TiOMagFig} The TiO(8442) spectral index versus I-band magnitude
for our targets. The plotting symbols used indicate binarity or membership of
the \sori\ or \lori\ clusters  or non-membership of these
clusters based on the radial velocity or spectral type (see
Fig.~\ref{EWTiOFig}), as indicated in the legend.  Error bars are only 
shown in cases where they are larger than the plotting symbol used.}
\end{figure}
\begin{figure}
\includegraphics[width=0.47\textwidth]{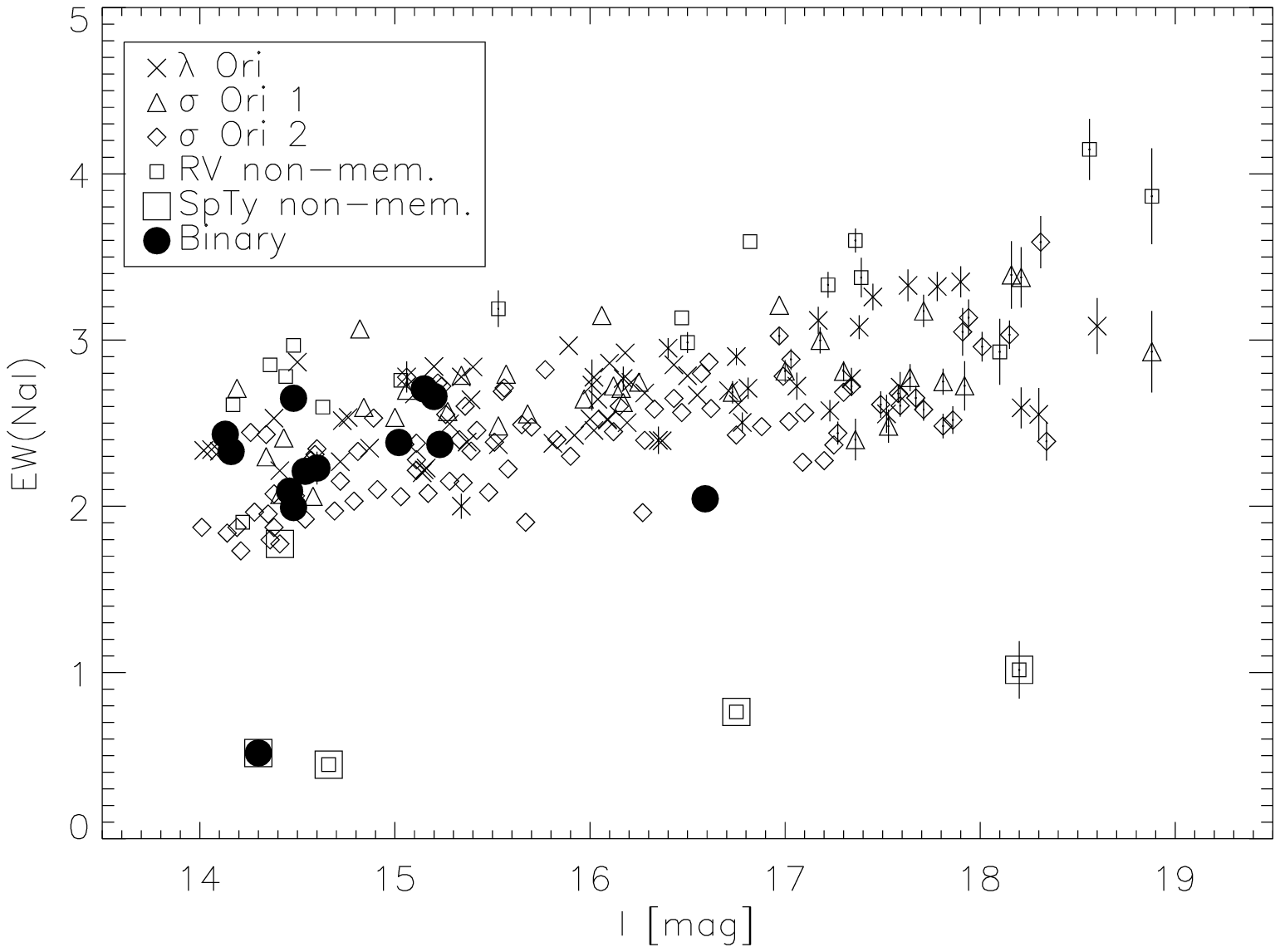}
\caption{\label{EWFig}The equivalent width of the Na\,I doublet for our
targets as a function of the I-band magnitude.  The plotting symbols used
indicate binarity or membership of the \sori\ or \lori\ clusters  or
non-membership of these clusters based on the radial velocity or spectral type
(see Fig.~\ref{EWTiOFig}), as indicated in the legend. Error bars are only 
shown in cases where they are larger than the plotting symbol used.}
\end{figure}
 In this section we describe the criteria we have used to identify members of
the \sori\ and \lori\ associations. The principal means of
identifying members of these clusters is the mean RV of the star,
but we have also used the equivalent width of the Na\,I doublet, EW(Na\,I),
and the strength of the TiO band at 8442\AA, TiO(8442), as additional
membership criteria.

 The distributions of weighted mean radial velocities,  \Vbar,  are shown
separately for stars near \sori\  and \lori\ in Fig.~\ref{RVDistFig}. The
bi-modal distribution of radial velocities for stars near \sori\
discussed by \citet{2006MNRAS.371L...6J} is apparent. We follow the convention
in that paper of assigning stars with   $\Vbar < 27$\,\kms\ to Group 1 and stars
with $\Vbar \ge 27$\,\kms\ to Group 2. The interpretation of these
groups favoured by \citeauthor{2006MNRAS.371L...6J} is that Group 1 are
members of either the Orion OB1a or OB1b association while Group 2 are a
separate cluster of stars associated with the star \sori. Group 2 are
concentrated spatially around the star \sori\ and have similar mean radial
velocity to it. Group 2 are younger on average than Group 1, although
considerable overlap is possible.

 For stars near \sori\ we identify non-members using the criterion
$\Vbar > 35\kms$ or $\Vbar < 20\kms$. For stars near  $\lambda$~Ori we
identify non-members using the criterion $\Vbar > 32\kms$ or $\Vbar  <
22\kms$. These limits are indicated in  Fig.~\ref{RVDistFig}. The application
of these criteria to stars with variable radial velocities are discussed in
more detail in section~\ref{BinaryNotes}. 

 The TiO bands around 8450\AA\ are good indicators of effective temperature
for M-dwarfs in the sense that they increase in strength for cooler stars
\citep{2004ApJ...609..854M} and are insensitive to reddening. We have measured
the strength of the band at 8442\AA\ in our spectra using the ratio of the
counts detected in the wavelength ranges 8437\,--\,8442\AA\ and
8442\,--\,8450\AA. The measurements were made on the median average spectrum
of each star. We denote the  value of this ratio as TiO(8442).
 
 The equivalent width of the Na\,I doublet at 8190\AA\ is sensitive to the
surface gravity in M-type stars \citep{1997ApJ...479..902S} in that stars with
higher surface gravities show stronger Na\,I absorption. The surface gravity
of members of the \sori\ and \lori\ clusters is expected to be $\logg =
3$\,--\,4, whereas a typical M-type giant will have $\logg \approx 1$. Thus,
the equivalent width of the Na\,I doublet, EW(Na\,I), can be used to identify
background giants in our sample. It is also possible to identify contamination
of the sample by dwarf stars with $\logg \goa 4.5$. We measured the value of
EW(Na\,I) from the median average spectrum of each star by numerically
integrating the area under our spectra in two regions $\pm 120\kms$ wide
around the centre of each Na\,I line after normalizing the spectra by the
clipped mean value of the spectrum in the region 8188\,--\,8192\,\AA.

The  value of  TiO(8442) is plotted against   EW(Na\,I) in
Fig.~\ref{EWTiOFig}. There is a clear division between the bulk of our targets
and the small group of stars with low EW(Na\,I) and TiO(8442) values. Most of
these stars are non-members based on their mean RV. We therefore
identify all stars below the dashed line  in this figure as ``spectral-type
non-members''.

From Figs.~\ref{TiOMagFig} and \ref{EWFig} we see that very few stars that
satisfy the RV  and spectral type criteria for membership have
discrepant values of EW(NaI) or  TiO(8442) for their \Ic\ magnitude. There are
two faint stars in the sample that have large values of TiO(8442), but there
is a large scatter in this index among the faint stars in our sample so we do
not consider this to be a reason to exclude these stars as members of the
\sori\ cluster. 

 The binary star with a discrepant value of EW(NaI) is J054052.5$-$021653.
This SB1 binary star has weak, narrow Na\,I absorption lines characteristic
of a giant star. The mean RV has a rather large uncertainty but
is also clearly inconsistent with membership of the \sori\ cluster. We
conclude that this is a background giant star. 

 The spectra of the star J054034.5$-$020606 extracted using our empirical
weighting scheme do not satisfy our criterion for inclusion in the sample
because the signal-to-noise is less than 5 but we did notice that this is an
SB2 spectroscopic binary from an analysis of the spectra extracted using
girBLDRS. These spectra are shown in Fig.~\ref{SpecFig}. This is the faintest
binary detected in our survey (\Ic = 18.38) so the spectra are quite noisy,
but the SB2 nature of this star can be seen and is very obvious in the CCF.
The values EW(Na\,I)$=3.42 \pm 0.23$ of TiO(8442)$ = 0.593 \pm 0.026$ shows
that this star has the wrong spectral type to be a member of the \sori\
cluster. We have also inspected the position of this star in the V~v.~V$-{\rm
I}_{\rm C}$ colour-magnitude diagram of the \sori\ region using the data of
\citet{2007MNRAS.375.1220M}, where it is clearly below the sequence of cluster
members. We conclude that this is a  background dwarf binary star for which
the combination of spectral type, distance and reddening place the star within
the band of cluster members in the \Ic~v.~$\Rc-\Ic$ colour-magnitude diagram.

 Having applied these membership selection criteria we find that in our sample 
there are 64 members of \lori, 34 members of \sori\ Group 1 and 86 members of
\sori\ Group 2, excluding stars with variable radial velocities that are
discussed separately below.

\begin{figure}
\includegraphics[width=0.47\textwidth]{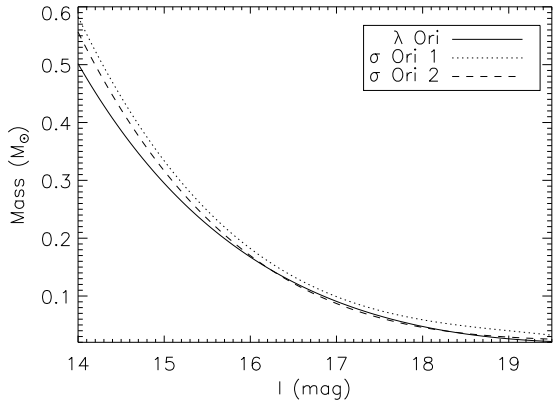}
\caption{\label{MagMassFig}The I-band mass-magnitude relations interpolated
from the models of \citet{1998A&A...337..403B} for the following distances and
ages: d=330pc, age=10\,Myr (\sori\ 1);  d=440\,pc, age=3\,Myr (\sori\ 2);
d=400\,pc, age=5\,Myr (\lori).
}
\end{figure}

\section{Binary stars in our sample}
\label{BinaryNotes}

 In this section we discuss the stars in our sample with variable radial
velocities or variable line profiles that are members of the
\sori\ or \lori\ clusters. The values of \Vbar, TiO(8442), etc. for these
stars are given in Table~\ref{BinaryTable}. The spectra of these stars in the
region of the Na\,I doublet are shown in Fig.~\ref{SpecFig}.

 We also note in Table~\ref{BinaryTable} whether only one set of spectral
lines are visible (SB1) or that the  spectra indicate the presence of two
stars (SB2). A star may be an SB1 binary either because the companion is too
faint to be detectable or because the lines of the stars are blended, or both.
We created a set of synthetic binary star spectra using combinations of
observed single star spectra over a range of luminosity ratio and radial
velocity difference to estimate the lower limit to the magnitude difference in
the SB1 stars. If we assume that neither star is rapidly rotating then we
estimate that any companion to these stars is more than 2.5 magnitudes fainter
than the primary. From the magnitude-mass relation shown in
Fig.~\ref{MagMassFig} we estimate that the corresponding limit to the mass
ratio is $q\loa 0.25$.

 In general, the \Ic\Rc\ photometry, the values of TiO(8442), EW(NaI) and the 
mean RV of these stars are all consistent with membership of
either the \lori\ cluster or one of the \sori\ clusters. We discuss exceptions
to this general rule and any other points of interest for these stars below.

\begin{description}
\item[\bf J053502.7$+$095649]{ \citet{2004ApJ...610.1064B} consider this star
(LOri-CFHT 043) to be a member of the \lori\ cluster on the basis of the
available R$_{\rm C}$I$_{\rm C}$JHK photometry. }
\item[\bf J053443.9$+$094835]{ \citet{2004ApJ...610.1064B} note that the
position of this star (LOri-CFHT 069) in the I$_{\rm C}$ v. I$_{\rm
C}-$K$_{\rm S}$ colour-magnitude diagram is inconsistent with cluster
membership. However, this is simply a result of the contribution of two stars
of similar brightness to the flux at I-band in this SB2 binary
(Fig.~\ref{SpecFig}). Correcting the I-band magnitude by 0.75\,magnitudes
places this star in a position entirely consistent with cluster membership in
this colour-magnitude diagram.}
\item[\bf J053455.2$+$100035]{ \citet{2004ApJ...610.1064B} note the presence
of H$\alpha$ emission in their low resolution  spectrum of this star
(LOri-CFHT 075). The position of this star in the I$_{\rm C}-$K$_{\rm S}$ v.
H$-$K$_{\rm S}$ colour-colour diagram and  the I$_{\rm C}$ v. I$_{\rm
C}-$K$_{\rm S}$ colour-magnitude diagram are also consistent with cluster
membership and with the spectral type of M5 assigned by
\citet{2004ApJ...610.1064B} so it is unclear why they note it as being a
non-member on the basis of this information in their Table 2.  The spectral
lines for this star show rotational broadening. We compared the spectra of
this star to those of a narrow-lined star of similar spectral type to which we
had applied a rotational broadening function for various values of the
projected rotational velocity,  $\Vsini$. From this comparison we estimate
that the projected rotational velocity of J053455.2$+$100035 is $\Vsini\approx
65\kms$. Only one set of spectral lines are visible in our spectra but there
is an asymmetry in the CCF in the form of a blue-wing, particularly when the
measured RV corresponds to a red-shift. The suggests that the
fainter component in this binary is detected but unresolved in the I-band
spectra. }
%
\item[\bf J053845.6$-$021157]{The range in radial velocities we have observed
for this star is $-6.9$ to $11.2\kms$. This is consistent with this star being
a member of \sori\ Group 1 if it has a semi-amplitude $\approx 30\kms$.
Members of Group 1 are approximately twice as common as members of Group 2 in
the 25\,arcmin {\sc giraffe} field used for the observations of this star.
The orbital period is required to be rather short ($P\loa 1$\,day) to
reconcile the observed radial velocities of this star with a mean velocity
consistent with cluster membership and the expected mass ratio for an SB1
binary ($q \loa 0.25$). }
\item[\bf J053801.0$-$024537]{The mean RV estimated from the spectra is
consistent with membership of either \sori\ Group 1 or \sori\ Group 2. There
are approximately 10 times as many members of Group 2 as Group 1 in this
field close to \sori\, so we conclude that it is approximately 10 times more
likely that this star is a member of Group 2 than Group 1. The values of
EW(Na\,I) and TiO(8442) for this star were measured from the spectrum observed
near conjunction. }
\item[\bf J054001.0$-$021959]{ \citet{2005MNRAS.356...89K} detected strong
lithium absorption in the spectrum of this star (KJN2005~6), which indicates
that this star is younger than 20Myr, as expected for a member of the \sori\
cluster. The mean RV of this star suggests it is a more likely to
be a member of \sori\ Group 2 than Group 1.}
\item[\bf J053823.5$-$024131]{ The mean RV measured from the
three spectra in which the two components are unresolved is
$26.5\pm0.9\kms$, which is close to the dividing line at 27\,\kms\ between
Group 1 and Group 2. The proximity of this star to \sori\ makes it
approximately 10 times more likely that this is a member of Group 2 than Group
1. This star (BMZ2001 S\,Ori J053823.6$-$024132) was listed as a candidate
member of the \sori\ cluster on the basis of IJHK photometry by
\citet{2004AN....325..705B}. }
\item[\bf J053743.5$-$020905]{We measured the value of EW(Na\,I) and TiO(8442)
for this SB2 binary from the spectrum taken near conjunction, so the values
represent an average value for the two stars, which are of comparable
brightness.  The RV measured from this spectrum is 42\,\kms,
which is outside the range we have defined for membership of the \sori\
cluster, but it is difficult to establish whether this is an accurate estimate
of the mean RV of this star. Further observations will be
required to establish the true mean RV of this binary star in
order to check that it is consistent with this star being a cluster
member. Members of Group 1 are approximately twice as common in this field as
Group 2 \citep{2006MNRAS.371L...6J}. For the purposes of this paper we assume
that this star is a cluster member. }
\item[\bf J053838.1$-$023202]{The mean RV of this SB1 binary and
its proximity to \sori\ suggest that it is a member of Group 2.  }
\item[\bf J053805.6$-$024019]{ The range of radial velocities observed in this
SB1 binary star is 4.4 to 14.5\kms, which is consistent with membership of
\sori\ if the semi-amplitude of the spectroscopic orbit is $K\approx 20\kms$.
The proximity of this star to \sori\ makes it approximately 10 times more
likely that this is a member of Group 2 than Group 1. }
 \end{description}

 In summary, we have identified 5 spectroscopic binary members of the \lori\
cluster and 7 spectroscopic binary members of the \sori\ cluster. Among the
\sori\ spectroscopic binaries, 5 stars are likely to be members of Group 2, 1
is likely to be a member of Group 1 and 1 star cannot be assigned to either
group. Of the \Nbinmem\  spectroscopic binary cluster members, 6 are SB2
binaries with stars of comparable brightness in the I-band. There is  1 SB1
binary with broad lines due to rotation so it is not clear whether this is a
genuine SB1 binary or an unresolved SB2 binary, although there is a hint of
the companion in the CCF for this star.
 There are 3 SB1 binaries with narrow lines for which we can
say the companion is likely to be more than 2.5\,magnitudes fainter than the
primary so that the mass ratio is less than about 0.25. We have also
identified two spectroscopic binaries that are not members of either \sori\
or \lori.
 
\subsection{The distribution of binaries with magnitude} 
It is notable that all 12 spectroscopic binary stars we have identified among
the \Nmember\ cluster members are brighter than $\Ic =16.6$, and 11 are brighter
than $\Ic = 15.25$.  There are 68 cluster members brighter than $\Ic =15.25$
including these binary stars. At face value, this suggests a close binary
fraction of about 16 percent above this limit and less than a few percent
below this limit. Of course, the sensitivity of our survey decreases for
fainter stars because the signal-to-noise of the spectra is less and the
orbital speeds decrease with mass for a given semi-major axis or orbital
period.

\begin{figure}
\includegraphics[width=0.47\textwidth]{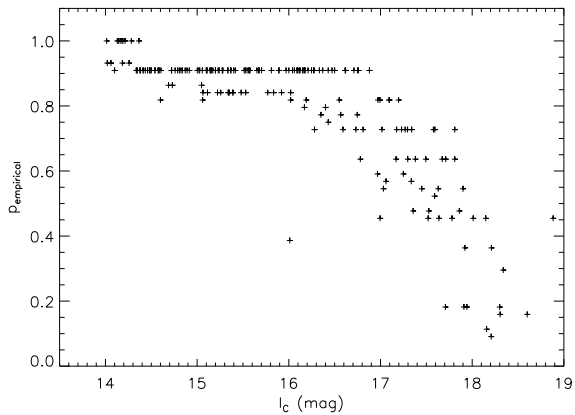}
\caption{\label{pdetect} The detection efficiency as a function of magnitude
for our survey based on the radial velocities of the binaries identified from
their radial velocity variations.}
\end{figure}

 We have used the RV data for the 11 binary cluster members we detected from
their RV variations to estimate the sensitivity of our survey to binaries of
this type as a function of I-band magnitude, $p_{\rm empirical}$. We exclude
J054001.0$-$021959  from this analysis because it is unclear how to simulate
the method by which we detected this  binary, i.e., from the variation of the
width of its CCF. For every combination of  variable and non-variable star in
our survey we have created a synthetic RV dataset using the RV errors in the
non-variable star dataset and the radial velocities of the binary star scaled
as follows. We first subtract our best estimate of the systemic radial
velocity for the binary. We then estimate the total mass of the binary,
$m_{\rm T}$ assuming that the stars are identical for the SB2 binaries or that
the companion is 2.5 magnitudes fainter than the primary for the SB1 binaries.
The masses are estimated using the models of \citet{1998A&A...337..403B} based
on the \Ic\ magnitude of the target. The relations between  \Ic\ and mass are
shown in Fig.~\ref{MagMassFig}. We have assumed the following  distances and
ages for the clusters: d=330pc, age=10\,Myr (\sori\ 1);  d=440\,pc, age=3\,Myr
(\sori\ 2); d=400\,pc, age=5\,Myr (\lori).  We then repeat the calculation to
find the total mass, $M_{\rm T}$, of a similar binary star with the same \Ic\
magnitude as the single star, $m_S$. We then multiply the radial velocities by
$\sqrt{m_S/m_{\rm T}}$. This is equivalent to assuming that the distribution
of semi-major axis is independent of mass. We then apply the same detection
criterion as before ($\log(p) < -4$) to the synthetic datasets.  In cases
where $N_{\rm rv}$ is less for the single star than the binary star we use the
average number of detections for all combinations of  $N_{\rm rv}$ synthetic
radial velocities from the combinations available. For cases where $N_{\rm
rv}$ is larger for the single star than the binary star we use the $N_{\rm
rv}$ measurements with the lowest RV errors to create the synthetic dataset.
The number of synthetic RV data sets which satisfy our variability criterion
then gives an estimate of the detection efficiency for each star, e.g., if 6
of the 11 synthetic RV data sets for a star satisfy our variability criterion,
then the same observations of a binary star similar to those discovered in
this survey with the same \Ic\ magnitude would, on average, have detected a
significant RV shift about 55\,percent of the time. The values of $p_{\rm
empirical}$ calculated in this way is shown in Fig.~\ref{pdetect} as a
function of the \Ic\ magnitude.

\begin{figure}
\includegraphics[width=0.47\textwidth]{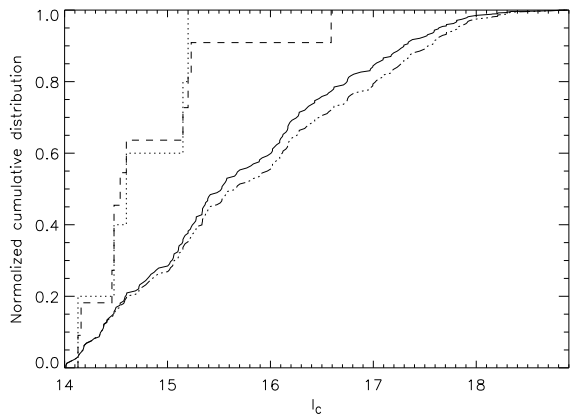}
\caption{\label{KSFig} The  normalized cumulative distribution of
the detection efficiency of our sample as a function of \Ic\ magnitude based
on the measured radial velocities of all  binaries detected (solid line)
or the SB2 binaries only (dashed-dotted line) compared to the normalized
cumulative distribution of \Ic\ magnitude for all the binaries discovered by
our survey (dashed line) and the SB2 binaries discovered by our survey (dotted
line).}
\end{figure}

 The normalized cumulative distribution function for these detection
efficiencies is shown as a function of \Ic\ magnitude in Fig.~\ref{KSFig}.
Also shown is the  cumulative distribution function for the \Ic\ magnitude of
the binary cluster members excluding J054001.0$-$021959. If the binary
fraction and semi-major axis distribution for binaries is independent of mass
then these two distribution should be the same. It is clear that the two
distributions are not the same. The  Kolmogorov-Smirnov test  applied to these
distributions gives a 99.7 percent significance level to the difference in
these distributions. There is a chance that a few of the stars we have
identified as SB1 binary stars are the result of spurious RV shifts due to
errors in the analysis or instrumental effects or the result of intrinsic
variability of a single star. If we take a very cautious approach and repeat
this analysis using the detections of SB2 binaries only we find that
Kolmogorov-Smirnov test gives significance level of 97.6 percent.

\subsection{The binary fraction\label{BinFracSec}} 

 We used a Monte-Carlo simulation to calculate probability distribution for
the binary fraction given the 11 binaries we have discovered from their
variable radial velocities and the detection efficiency for each star, $p_{\rm
empirical}$, calculated above.  

\begin{figure}
\includegraphics[width=0.47\textwidth]{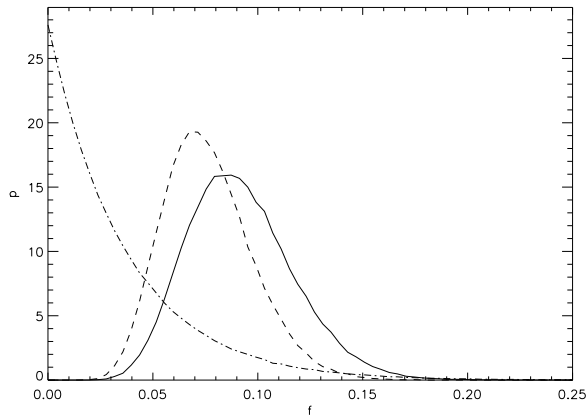}
\caption{\label{BinFracFig} The probability distribution for
the binary fraction in our whole sample (dashed line), the `bright' sample
(solid lines, mass $> 0.1\Msolar$) and the `faint' sample 
(dashed-dotted line, mass $<0.1\Msolar$). }
\end{figure}

 The results are shown in Fig.~\ref{BinFracFig} for the whole sample, a
`bright' sample ($\Ic < 16.9$) and a `faint' sample ($\Ic \ge 16.9$). The
division here between the bright and faint samples has been chosen to
correspond to the widely accepted division between low mass stars and VLM
stars at a mass of 0.1\Msolar \citep{2007prpl.conf..427B}. The mean value of
$p_{\rm empirical}$ for the 145 stars in the bright sample is 0.89. For the 51
stars in the faint sample the mean value of  $p_{\rm empirical}$  is 0.53. The
spectroscopic binary fraction for the bright sample is $ 9.5^{+1.2}_{-2.8}$
percent. The 90 percent confidence upper limit to the spectroscopic binary
fraction for the faint sample, given the assumptions above, is 7.5 percent.
These figures apply to binary stars of the type discovered by our survey only,
i.e,. we have not attempted to apply a correction to this spectroscopic
 binary fraction for the binaries at longer orbital periods that are not
detected by our survey. The hypothesis that $f_{\rm bright}$ and $f_{\rm
faint}$  are equal can be rejected with 90\,percent confidence. The value of
$\Ic=16.9$ that is used as the dividing line between the faint and bright
samples is arbitrary. It would be possible to achieve a higher level of
significance by setting the limit between our bright and faint samples closer
to magnitude of the faintest binary in our sample ($\Ic=16.5$), but this would
be an equally arbitrary value and so it would be hard to justify the apparent
increase in statistical significance.

\subsection{The detection efficiency as a function of binary separation}

\label{method}

 We have used another Monte Carlo simulation to estimate the range of
semi-major axis, $a$,  over which binary stars  can be detected by our survey.

 If binarity is the only cause of variable RVs, the probability that a given
target is flagged as an RV variable is given by $\epsilon_b p_{\rm
detect}+(1-\epsilon_b)10^{-4}$, where $\epsilon_b$ is the overall binary
fraction and $p_{\rm detect}$ is the probability that $\log p < -4$  for the
object assuming that it is a binary.

 We have used a Monte Carlo simulation to calculate the value of $p_{\rm
detect}$ as a function of semi-major axis, $a$, for every star in our sample
given various assumptions about the distribution of binary properties. The
simulation generates 65536 virtual binaries  and predicts the RV of the more
massive component at the same times of observation as the actual observations.
The eccentricity, $e$, mass ratio, $q$, and other properties of the binary
star are randomly selected from the following distributions.

\begin{description}
\item[\bf Mass ratio, {\boldmath $q$}]{We have used  a `flat' distribution
which is uniform in the range $q=0.2$--1. We did not consider the peaked mass
ratio distribution we investigated in \citet{2005MNRAS.362L..45M} to be
appropriate for the binaries we have discovered in this survey because that
distribution is zero for $q<0.7$ whereas some of the SB1 binaries we have
found must have mass ratios $q\loa 0.25$. The value of the mass ratio makes
little difference to the value of $p_{\rm detect}$  in this range.}
\item[\bf Eccentricity, {\boldmath $e$}]{We have assumed that all binaries
with periods less than 10\,d have circular orbits \citep{2005ApJ...620..970M}.
Above this period, we assume that the value of $e$ is uniformly distributed in
the range $e=0$--$e_{\rm max}$ where $e_{\rm max}= 0.6$. We have also
performed one set of simulations with $e_{\rm max}=0$ in order to investigate
the effect of assuming that all orbits are  circular.}
\item[\bf Primary mass,  {\boldmath $m$}]{We have calculated the mass of the
primary star based on its \Ic\ magnitude and the models of
\citet{1998A&A...337..403B}. The mass-magnitude relations shown in
Fig.~\ref{MagMassFig} are  used to find a primary star mass consistent with
the observed magnitude and the mass ratio of the synthetic binary. We allow
for an assumed error of 0.03\,magnitudes in the observed \Ic\ magnitude. We
find that the choice of mass-magnitude relation for each star has a negligible
affect on our results. For simplicity, we present the results assuming a
distance of 330\,parsec and an age of 10\,Myr for all stars.}
\item[\bf Orbital phase]{The orbital phase of the binary  at the date of the
first observation is randomly selected from a uniform distribution in the
range 0 to 1.}
\item[\bf Longitude of periastron, {\boldmath $\omega$}]{For eccentric
binaries, $\omega$ is selected from a  uniform distribution in the range 0 to
$2\pi$.} \end{description}

 We have calibrated the extent to which blending between the components
reduces the apparent amplitude of the RV variation in a
spectroscopic binary. We selected 6 stars with a range of \Ic\ magnitudes that
had typical spectra for stars of that magnitude. For each pair of stars we
created simulated binary star spectra in which the single star spectra were
combined with the appropriate flux ratio for their magnitude difference and a
range of velocity offsets between the stars. We then measured the radial
velocity of the brighter star in the spectrum in the same way as we did for
our actual observations. We used these results to calibrate the difference
between the true RV and the apparent RV of the
brighter star caused by blending with the spectrum of the fainter star. We
used interpolation within the resulting table to adjust the RV of
the more massive star in each simulated binary star to account for this
blending.

 The radial velocities predicted by each trial of the simulation are each
perturbed by a random value from a Gaussian distribution with the same
standard deviation as the random error of the actual observations. We then
estimated the range of inclinations over which the binary would be detected
using the same criterion that we applied to our actual data. In the absence of
blending this is a trivial calculation. In the presence of blending the value
of $\chi^2$ may not be a monotonic function of inclination, $i$. We
approximate the true shape of the relation between $\chi^2$ and $\sin i$ using
a parabolic fit to the values at $\sin i = 0, 0.5$ and 1 and use this
parabolic fit to estimate the range of $\sin i$ values over which the binary
would be detected. 

\begin{figure}
\includegraphics[width=0.47\textwidth]{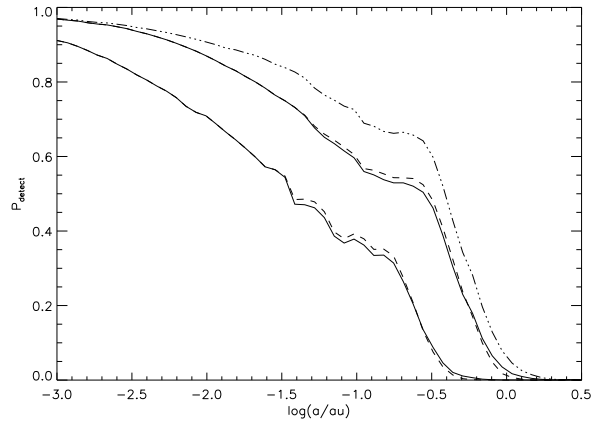}
\caption{\label{efficiency} The average detection efficiency for our survey as
a function of semi-major axis ($a$) for the bright sample (upper curves) and
the faint sample (lower curves). The solid line corresponds to $e_{\rm max} =
0.6$,  the dashed line to $e_{\rm max} = 0$ and the dashed-dotted line shows
the effect of neglecting blending in the case $e_{\rm max} = 0.6$ for the bright
sample.}
\end{figure}

 The average value of $p_{\rm detect}$ calculated in this way for all the
stars in the bright and faint samples are shown as a function of $\log a$ in
Fig.~\ref{efficiency}. From that figure we see that assuming circular orbits
does not make a large difference to the value of $p_{\rm detect}$, but that
neglecting blending can lead to an overestimate of the detection efficiency by
as much as 15 percent. We can also see that the detection efficiencies
calculated above using the RV data of the binaries themselves gives a result
consistent with the values of  $p_{\rm detect}$ in Fig.~\ref{efficiency} if
the semi-major axes of these binaries are in the range  $-2.5 \loa \log (a/au)
\loa -1.0$.  The corresponding range in orbital periods is $ 0.1\,{\rm d} \loa
P \loa 30\,{\rm d}$, which is in good agreement with the likely orbital
periods of the binaries we have detected.

\subsection{Comparison with \citet{2005MNRAS.356...89K} \label{kenyon}}
 We combined our radial velocity data with those of
\citet{2005MNRAS.356...89K} to see if the combined data sets would yield any
further spectroscopic binaries. We did not find any new binaries among the 45
stars in common between the two surveys. This is, perhaps, not surprising
given the much higher radial velocity accuracy of our data compared to
\citeauthor{2005MNRAS.356...89K}. This is as a result of the higher
signal-to-noise, higher resolution and superior telluric subtraction
achievable with {\sc giraffe} spectrograph compared with the {\sc wyffos}
spectra available to Kenyon et~al. The mean difference in the measured radial
velocity between the two sets of data is $0.72\pm 0.44\,\kms$. 

 The star KJN2005~72 (J053739.6$-$021826) was identified as a possible
spectroscopic binary by \citet{2005MNRAS.356...89K} based on a RV
shift of $35 \pm 4$\,\kms\ between two spectra obtained on consecutive nights.
The radial velocities for this star measured from our data are constant to
within a 2 \kms\ over a baseline of 63 days.  We conclude that the radial
velocity shift measured by \citet{2005MNRAS.356...89K} for this star is likely
to be spurious.

Similarly, \citeauthor{2005MNRAS.356...89K} claim a RV shift of
$13\pm 6$\,\kms\ for the star KJN2005~74 (J053926.8$-$023656) between two
spectra obtained on consecutive nights. We find the RV for this
star is constant to within 2\kms\ from 3 spectra with a baseline of 61 days.

\citeauthor{2005MNRAS.356...89K}  note that the star  KJN2005~46
(J054000.1$-$025159) appears to be a member based on the presence of the Li\,I
6707\AA\ line in the spectrum and the equivalent width of the Na\,I doublet,
but the RV they measure ($17\pm 3$\,\kms) is inconsistent with
cluster membership. They suggest that this may be due to this star being
spectroscopic binary. However, we find that this star has a mean radial
velocity of 30.5\kms\ which is consistent with cluster membership and is
constant to within 1\kms\ from 3 spectra with a baseline of 61 days.

In summary, it appears that the radial velocities measured by
\citeauthor{2005MNRAS.356...89K} sometimes show spurious shifts of
approximately 10\kms. 

\section{Discussion}

 There are several examples of VLMS and BDs that are clearly spectroscopic
binaries. \citet{2006Natur.440..311S} measured accurate masses and radii for
the brown dwarf pair 2MASS J05352184$-$0546085  which is an eclipsing
spectroscopic binary with an orbital period of 9.8\,days and a total mass of
0.088\Msolar\ in the Orion Nebula cluster (ONC). PPl~15 in the Pleiades was the
first brown dwarf confirmed by the detection of lithium and is an SB2
spectroscopic binary containing two brown dwarfs with masses of 60\,--\,70
Jupiter masses,  an orbital period of 5.8\,days and an eccentricity of 0.4
\citep{ 1999AJ....118.2460B}. There is no large RV survey for
spectroscopic  binary VLMS and BDs in the Pleiades so we do not know whether
PPl~15 is representative of the binary population in this cluster. Similar
arguments apply to 2MASS J05352184$-$0546085 and the ONC. Intriguingly, there
is a very well populated binary sequence in the colour-magnitude diagram for
the Pleiades \citep{2007arXiv0706.2234L}, which suggests a binary frequency of
28\,--\,44 percent in the 0.075\,--\,0.030\Msolar\ mass range.  

Surveys for spectroscopic binaries among late-M and T dwarfs in the field and
in nearby clusters have discovered a few other SB2 binaries and  several stars
that show RV shifts of a few \kms\ or less.
\citet{2006AJ....132..663B} summarize the results of these surveys and present
the results of their own survey of 53 VLMS and BDs. From their own sample they
estimate a binary frequency of 11 percent in the separation range 0\,--\,6\,au.
This binary frequency may be consistent with our results for VLM binaries
(binary fraction $<7$ percent for mass $\loa  0.1$\Msolar) if the
distribution of $a$ is not strongly biased towards small $a$. This is
reasonable given that none of the binaries detected by
\citeauthor{2006AJ....132..663B} were SB2 binaries and that the typical radial
velocity shifts detected were small (few \kms).

 It is much harder to interpret the results of previous surveys summarized by
\citet{2006AJ....132..663B}, not only because of the small number of stars
observed but also because there is a bias in these surveys due to
preferentially observing brighter stars or imposing a magnitude limit. This
tends to increase the number of binaries in these samples, particularly SB2
binaries. Our survey is much less strongly affected by this type of bias. Some
SB2 stars will be missing from our sample because they exceed our bright
magnitude cut-off. The lack of faint binaries in our sample means there is no
compensating gain in SB2 binaries at the faint end of the sample. This bias
can only increase the statistical significance of the change in binary
properties near $\Ic\approx 16.9$ that we have discovered. It is difficult to
estimate the compensating bias introduced by excluding stars below the cluster
sequence in the \Ic~v.~$\Rc-\Ic$ colour magnitude diagram but this effect is
expected to be small. 

 It is also harder to characterize the sources of noise in small surveys. This
is apparent from the results of \citet{2005MNRAS.356...89K} in which the few
RV shifts they measured of about 10\,\kms\  were ascribed to
binary motion, both by \citeauthor{2005MNRAS.356...89K}  and by
\citet{2005MNRAS.362L..45M}. In that case it appears that there is some
problem with obtaining reliable  RV measurements from spectra
affected by telluric absorption with low resolution and low
signal-to-noise. A similar problem applies to the interpretation of radial
velocity shifts $\loa 1\,\kms$, even for spectra of the highest quality. At
this level, line profile variations  due to chromospheric activity and star
spots (``jitter'') may become important, but the extent of this effect has not
been  well characterized. Our survey may also suffer from this problem in the
case of some of the SB1 binaries we have detected. For stars with narrow lines
and large RV shifts such as J053805.6$-$024019 this is unlikely
to be an issue. For example, \citet{2007arXiv0710.2437J} have shown that the
star \ChaHa\  is a binary star with an eccentric orbit ($e=0.49$) and a low
mass companion in a wide orbit ($a\approx$ 1 au). The semi-amplitude of the
orbit is low ($K=1.6\kms$) but can be measured reliably because the level of
jitter in this  young, narrow-lined VLM star is much less than $ 1\,\kms$. It
is less certain that stars like J053455.2$+$100035 with broad spectral lines
and small RV shifts are genuine spectroscopic binaries. One
mitigating factor in our survey is that there are several other stars in our
survey with broad lines that are not detected as binary stars, so it does
appear that we can measure radial velocities accurate to about 1\kms\ in this
type of star. Nevertheless, it remains to be established that any of the
single-lined stars showing RV shifts from this survey, the survey
by \citet{2006AJ....132..663B} or the survey by \citealt{2003A&A...401..677G}
are genuine SB1 spectroscopic binaries with Keplerian orbits.

Note, however, that the statistical significance of the change in binary
properties for very low mass stars in this survey we have noted  remains high
($>97$ percent) even if we consider only the SB2 binaries. 

  It is harder to detect binary stars among rapidly rotating stars because 
the RV measurements from  broader CCFs is less precise than for
narrow CCFs. This does not affect our conclusion that there is a  change in
binary properties for very low mass stars compared to low-mass stars because
rapidly rotating stars are evenly distributed in magnitude within our sample
and the standard errors of the radial velocities we assign to these stars
accurately accounts for the effects of line broadening.

 A rapid change in close binary fraction at a given magnitude is
only approximation to what is, in reality, likely to be a gradual change in
binary properties with mass. We do not have enough data to be able to say with
any accuracy at which mass this change occurs or whether the change occurs
over a small or large range of masses. This is why we have presented results
with the sample divided into  bright and faint sub-samples at $\Ic=16.9$, a
magnitude that corresponds to the commonly accepted dividing line between
VLMS/BDs and low-mass stars at 0.1\Msolar.
 
 With only 11 binaries and 3 or 4 RV measurements per star we
cannot say a great deal about the distribution of the binaries properties in
low mass star with the data available so far. However, we do have estimates of
the mass ratio in these binaries. The mass ratio is clearly $q\approx 1$ in
the case of the SB2 binaries with nearly equal components (``twins''). For the
stars where no companion is visible and the spectral lines are narrow we
estimate that the companion is likely to be at least 2.5\,magnitudes fainter
than the primary and so $q\loa 0.25$. The star  J053455.2$+$100035 lies
somewhere between these extremes. The overall picture, then, is of a peak in
the mass ratio distribution near $q=1$ due to twins over-laid on a broader
distribution increasing towards small values of $q$. This is qualitatively
similar to the mass ratio distribution found for low mass Population~I stars by
\citet{2003ApJ...591..397G} and the mass ratio distribution for solar-type
stars measured by \citet{2003A&A...397..159H} for solar type stars in the
solar neighbourhood, Pleiades and Praesepe. \citeauthor{2003ApJ...591..397G}
find that the peak in the $q$ distribution due to twins is not present for
Pop~I binaries more massive than 0.67\Msolar\ or for halo stars.
\citeauthor{2003A&A...397..159H} find that the peak for $q > 0.8$ gradually
decreases when long-period binaries are considered. 
 
 We have estimated the number of binaries we would detect if the stars
brighter than \Ic=16.9 (mass $\approx 0.1\Msolar$) in our sample have the same
binary frequency and orbital period distribution as nearby M-dwarfs. We used
the detection efficiency for each star as a function of $\log P$ convolved
with the Gaussian distribution for $\log P$ from \citet{1991A&A...248..485D}
together with the binary fraction for M-dwarfs of 42\,percent from
\citet{1992ApJ...396..178F} to find that we would have detected 6.5
spectroscopic binaries an average in our survey given these assumptions. At
face value, this suggests that the frequency of short period binaries among
low mass stars  in \lori\ and \sori\ is slightly larger to that for nearby
M-dwarfs. However, the assumption that the distribution of $\log P$
established by \citeauthor{1991A&A...248..485D} for solar-type stars is
appropriate for low mass stars may not be correct.
\citeauthor{1992ApJ...396..178F} find that the period distribution for
M-dwarfs binaries is similar to that for solar type stars but there is only
one short period M-dwarf binary in the sample  so the period distribution is
poorly characterized at the short period end.
\citeauthor{1991A&A...248..485D} note that \citet{1985ib...proc....1G}
measured a much higher frequency of short period binaries for solar type stars
in the Hyades than is seen in their field star sample. The same effect was
seen to a lesser extent by \citeauthor{2003A&A...397..159H}, so both
environment and primary mass may be important factors in determining the
frequency of short period binaries in our sample. We do not have sufficient
data in our survey to  measure the orbital periods and semi-amplitudes of the
spectroscopic binaries we have detected. 

 Given the uncertainty in the orbital period distribution of low mass stars it
would clearly be worthwhile to obtain complete orbits for the binary stars we
have identified.  The maximum separations of the lines observed in the
SB2 binaries and the timescale and amplitude of the RV variations for the SB1
binary stars suggest that the orbital periods of these binaries are likely to
be in the range from several hours to several days. An anonymous referee has
raised the possibility that the difference in binary fraction we have observed
in the bright and faint samples is due to a distribution of $\log a$ strongly
biased towards  $\log(a/au) \goa -0.5$. It can be seen from
Fig.~\ref{efficiency} that we have almost no sensitivity to binaries within
the faint sample but good sensitivity to binaries within the bright sample in
this $\log a$ range. In principle, it may be possible to find a distribution
of $\log a$ that can explain the numbers of binaries detected in the bright
and faint samples for a single value of the binary fraction. However, such a
distribution would be incompatible with the observation that several of the
binaries we have discovered have short orbital periods corresponding to values
of $\log (a/au) \ll -0.5$. If it were the case that the binaries we have
detected are biased towards  $\log(a/au) \goa -0.5$ then the values of p$_{\rm
empirical}$ we calculated for stars in the faint sample would have been very
small. In this case, the probability distribution for the binary fraction
calculated in section~\ref{BinFracSec} would have been consistent with the
hypothesis  $f_{\rm faint} =  f_{\rm bright}$. In fact, this hypothesis can be
rejected with 90 percent confidence, so the binaries we have detected are not
biased towards  $\log(a/au) \goa -0.5$.
 
 The lack of binaries in our sample with masses $\loa 0.1$\Msolar\ is
consistent with the overall picture established from existing surveys that the
binary fraction for VLMS and BDs across the full range of orbital separations
is 20\,--\,25 percent (\citealt{2007ApJ...668..492A};
\citealt{2006AJ....132..663B}).  However, the binary properties for  VLMS and
BDs is poorly characterized in the range $a \loa 2\,{\rm au}$ that is below
the detection limits of high angular resolution imaging. Some VLMS and BDs
show apparent RV shifts comparable to the orbital speeds expected at these
separations, and a complete spectroscopic orbits has now  been published for
one such star (\ChaHa, \citealt{2007arXiv0710.2437J}). It is still possible
given all the observations to-date that there is a large population of VLMS/BD
binaries with $a\approx 1$\,au. If such a population exists then it may be
possible to reconcile the binary fraction in existing surveys with the high
binary fraction inferred for the Pleiades by \citet{2007arXiv0706.2234L}.

 The spectra we have used for this study were obtained with the aim of
measuring the binary properties of the low mass stars and brown dwarfs in the
\lori\ and \sori\ clusters but there is clearly useful information to be
obtained about the properties of the clusters themselves from the spectra we
have obtained. For example, the distribution of EW(Na\,I) shown in
Fig.~\ref{EWFig} clearly shows that the stars in \lori\ have higher surface
gravities on average than stars in Group 2 of the \sori\ cluster. This shows
that the \lori\ cluster is older than the Group 2 of the \sori\ cluster and,
perhaps, has a similar age to Group 1 of the \sori\ cluster. This is in
general agreement with the ages we have adopted for these populations. A more
detailed analysis is beyond the scope of this paper. 

\section{Conclusions}

 We have conducted a large radial velocity survey for spectroscopic binary
stars among low mass stars and brown dwarfs in the young clusters around
\lori\ and  \sori. We have identified \Nmember\ members of these clusters
based on their radial velocity and spectral type. Of these, 6 are SB2 binaries
and 6 are SB1 binaries. All the spectroscopic binaries we have detected are
brighter than \Ic=16.6  (mass $\approx$ 0.12\Msolar). We conclude that the
frequency of spectroscopic binaries in these clusters among very low mass
stars (mass $<0.1\Msolar$) and brown dwarfs is significantly lower ($<7.5$
percent) than that for more massive stars ($9\pm 2$ percent). The change in
binary properties with mass that we have discovered may be due to a change in
the total binary frequency with mass  or a change in the period distribution
of the binaries with mass or both.

 The number of SB2 binaries in this sample suggests there may be a peak in the
mass ratio distribution for spectroscopic binaries in these clusters near
$q\approx 1$. There is also clear evidence from the properties of the SB1
binaries that the mass ratio distribution for spectroscopic binaries in these
clusters is broad, extending down to $q\approx 0.25$.

\section*{Acknowledgments}
 Based on observations collected at the European Southern Observatory, Chile
(Programme ID: 076.C\_145). RJJ was supported by a Nuffield Undergraduate
Research Bursary. This research is partially funded by a Science \& Technology
Facilities Council research grant (formerly PPARC). We thank the  referee for
comments that helped to improve the clarity of this paper.

\label{lastpage}
\bibliographystyle{mn2e}  
\bibliography{mybib}

\end{document}